\documentclass[a4paper,11pt]{article}
\pdfoutput=1 

\usepackage{jcappub} 

\usepackage[T1]{fontenc} 
\usepackage{graphicx}
\usepackage{dcolumn}
\usepackage{bm}
\usepackage{physics}
\usepackage{hyperref}
\usepackage[mathlines]{lineno}
\usepackage{youngtab}
\usepackage{natbib}
\usepackage[utf8]{inputenc}
\usepackage{slashed}
\usepackage{caption}
\usepackage{subcaption}

\newcommand{\gt}{\tilde{g}}
\newcommand{\deuteron}{D}
\newcommand{\cmcs}{\textrm{cm}^3/\textrm{s}}
\newcommand{\dash}{\text{--}}

\newcommand{\tuniv}{t_{\textrm{univ}}}

\usepackage{soul}

\defcitealias{FarrarZaharijas}{FZ03}
\def\FZ{\citetalias{FarrarZaharijas}}

\title{\boldmath Constraints on long-lived di-baryons and di-baryonic dark matter}

 \author{Glennys R. Farrar}
 \author{and Zihui Wang}
 \affiliation{Center for Cosmology and Particle Physics, Department of Physics, New York University, New York, NY 10003, USA}

\emailAdd{gf25@nyu.edu}
\emailAdd{zihui.wang@nyu.edu}

\abstract{A color-flavor-spin singlet state of six quarks $uuddss$ ($S$, or sexaquark) has been argued to be a potentially undiscovered deeply bound, long-lived hadron. Theoretical calculations of the $S$ mass have been made in the literature, widely varying from deeply-bound $\sim 1.2$ GeV to weakly-bound $\sim2.2$ GeV. Given the spread of the mass predictions, it is vital to derive observational constraints on the state as a function of the mass. The transition rates between $S$ and two baryons are governed by $\tilde{g}$, the effective Yukawa coupling between $S$ and two baryons with the same quantum numbers as $S$. In this paper, we place strong observational constraints on $\tilde{g}$, improving on various previous limits; additional limits assuming $S$ is dark matter are also presented. }

\begin{document}
\maketitle
\flushbottom

\section{Introduction}
\label{sec:intro}

The possibility that a deeply-bound $uuddss$ sexaquark ($S$) with two units of baryon number may be an undiscovered stable hadron is advocated in~\cite{Farrar17,Farrar:2022mih}. 
Unlike the case for bound states of two flavor octet baryons, Fermi statistics does not prevent all 6 quarks from being in a fully symmetric spatial configuration.  Moreover, as a singlet simultaneously under color, spin and flavor, the color magnetic binding between quarks is maximal because the relevant quadratic Casimir operators vanish, as pointed out by Jaffe~\cite{Jaffe77}.  Using the MIT bag model, Jaffe calculated the mass of the lightest such state, the H-dibaryon, to be 2150 MeV; this mass is high enough to allow singly-weak decay, giving the H-dibaryon a lifetime estimated to be of order $10^{-10}$ s~\cite{Jaffe77} but later calculated to be of order $10^{-8}$ s~\cite{donoghue86}. Subsequent mass predictions using a variety of models gave results spanning from 1.2 GeV to heavier than 2.2 GeV, thus leaving open the possibility of a deeply bound state with a cosmologically long lifetime in the event of a mass less than $m_\Lambda + m_n$~\cite{FarrarZaharijas} or absolutely stable if it is lighter than two nucleons. The mass estimates are reviewed in greater detail in Sec.~\ref{sec:pheno}. 

The deeply-bound $S$, if it exists, is distinct from and completely orthogonal to a weakly bound H-dibaryon composed of $uuddss$. Experiments creating doubly-strange hypernuclei and observing the decay products of $\Lambda$'s~\cite{hyper1991,hyper20011,hyper20012,hyper2010}, showed that the formation time of an H is longer than the $\Lambda$ lifetime.  This would not be the case for a loosely bound, ``molecular'' H-dibaryon, unless its binding energy is less than the binding energy of two $\Lambda$'s in the hypernucleus, where $B_{\Lambda\Lambda} \approx 7$ MeV in ${}^6_{\Lambda\Lambda}\mathrm{He}$~\cite{hyper20012}). Thus these experiments exclude the existence of a weakly-bound $uuddss$ state with mass between $m_n + m_\Lambda = 2054$ MeV and 2223.7 MeV~\cite{hyper20012}.  The latest lattice QCD calculations~\cite{SasakiHALQCD20,greenMainzH21} show either a very weakly bound $\Lambda \Lambda$ molecule or above-threshold resonance, consistent with the hypernuclear experiments and an ALICE $\Lambda \Lambda$ phase-shift analysis~\cite{ALICE_LamLam19}.  In the following, we are not concerned with the H-dibaryon, but rather we constrain  a possible compact, spatially symmetric, deeply bound state of the same quarks, designated $S$.

\begin{figure}[htb]
    \centering
\includegraphics[width=0.25\textwidth]{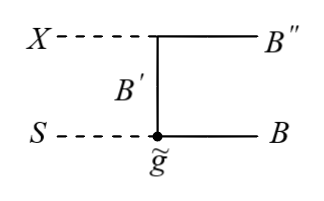}
    \caption{The effective Feynman diagram for $XS\to BB''$. The lower vertex, $\gt$, is an effective Yukawa vertex for $S$ dissociating to two baryons that have the same quantum numbers as $S$ (e.g. $\Lambda\Lambda$, $\Sigma^+\Sigma^-$ or $\Xi N$).}
    \label{fig:int}
\end{figure}

Ref.~\cite{Farrar18} studied the formation of $S$ at the QCD phase transition in the early Universe using statistical mechanics and found that the abundance of $S$ is comparable to that of ordinary baryons, and that with a sufficiently small breakup cross section the sexaquark component of the baryon asymmetry could survive the hot hadronic phase. Studies by ~\cite{Strumia18,KolbTurner18}, assumed that $S$ would thermalize with ordinary baryons in the hot hadronic phase of the early Universe and obtained the thermal evolution of $S$. Kolb and Turner~\cite{KolbTurner18}, taking breakup reactions $XS \leftrightarrow BB'$\footnote{Examples are $K S \leftrightarrow p\Lambda $, $\pi S \leftrightarrow \Sigma \Lambda$, $\gamma S \leftrightarrow \Lambda \Lambda $ and $\pi\pi S \leftrightarrow \Lambda \Lambda$. Note that $\pi S \leftrightarrow \Lambda \Lambda$ is forbidden by isospin conservation.} to have a typical hadronic cross section $\sim 1/m_\pi^2$, concluded that the density of $S$ today is at most $10^{-10}$ that of ordinary baryons but if $S$ exists it is an interesting relic. It is however worth noting that in Ref.~\cite{KolbTurner18} Kolb and Turner noted two scenarios in which $S$ can have a significant dark matter (DM) relic density. In the first scenario, $S$ must be lighter than 1.2 GeV, in accordance with~\cite{Strumia18}.  However as we show in Sec.~\ref{sec:pheno}, this light mass scenario is not viable because it leads to the transition of two bound nucleons in a nucleus into $S$ and a kaon with singly-weak lifetime, and is therefore ruled out by the stability of nuclei.  In the second scenario, the one advanced in Ref.~\cite{Farrar18}, the $S$ breakup cross sections must be orders of magnitude smaller than $1/m_\pi^2$ to prevent $S$ from coming into chemical equilibrium with baryons. Ref.~\cite{Farrar18} argued that such small breakup cross sections can arise naturally because the effective Yukawa coupling of $S$ with two baryons that have identical quantum numbers as $S$ (such as $\Lambda \Lambda$), denoted $\gt$ may be very small. Details of that argument are given in appendix~\ref{app:relic}. The generic Feynman diagram for $XS\to BB''$ including the Yukawa vertex is shown in figure~\ref{fig:int}.

Given the wide variance in $m_S$ prediction using different theoretical calculation techniques (1.2 GeV - 2.23 GeV) and the possibility that $S$ could be produced in the early Universe, we are motivated to derive observational constraints on $\gt$ in the entire mass window 1.2 GeV - 2.23 GeV. The potential that a sufficiently small $\gt$ may permit a relic abundance of $S$ sufficient to compose up to 100\% DM adds further motivation to study $\gt$. The Yukawa coupling $\gt$ governs the transition between $S$ and two baryons and produces a rich landscape of phenomenology. For instance, if $m_S$ is lighter than two bound nucleons, a nucleus could decay to $S$ and the rate is governed by $\gt$. We discuss under what circumstances the $S$ is consistent with constraints on the stability of nuclei. If $S$ is furthermore to be the DM, the possibility of its decay to baryons throughout cosmic history generates observable effects and we derive additional limits on $\gt$ based on this consideration  It should be stressed that the scattering cross section between $S$ and baryons is expected to be of order mb~\cite{Farrar17,Farrar:2022mih} and is not related to $\gt$.

Our paper is organized as follows. In Sec.~\ref{sec:pheno}, we overview the phenomenology of $S$ in different regimes of $m_S$ and present the main result of the paper --- the excluded region of $\gt$ as a function of $m_S$. Section~\ref{sec:nuclei}--\ref{sec:Hdibaryon} are dedicated to deriving the various observational limits on $\gt$; in Sec.~\ref{sec:nuclei} we constrain the decay of the deuteron to $S$ and obtain strong limits from the Sudbury Neutrino Observatory (SNO) experiment; in Sec.~\ref{sec:S}, we analyze the signatures of $S$ DM decaying to two baryons; in Sec.~\ref{sec:Hdibaryon}, we review the previous searches for the H-dibaryon and assess their sensitivity to $\gt$. In Sec.~\ref{sec:BBG} we discuss a possible model that predicts a small $\gt$. Finally, we end with a review of existing literature on $S$ in Sec.~\ref{sec:critics}.

\section{Landscape of \texorpdfstring{$S$}{S} phenomenology}
\label{sec:pheno}

\subsection{Interactions of \texorpdfstring{$S$}{S}}
Due to baryon number conservation, there are only two types of effective stong-interaction vertices of $S$ on the hadron level. The first type is a Yukawa vertex that transfers baryon number between two baryons, $B$ and $B'$, and $S$, where the two baryons have identical net quantum numbers as $S$. The other type is a vertex where $S$ emits or absorbs a flavor singlet meson, predominately a vector meson $V^\mu$.\footnote{Since the $S$ is a singlet in flavor, it cannot emit a flavor-octet meson without transitioning to a much more massive non-singlet di-baryon. The flavor-singlet combination of the $\omega$ and $\phi$~\cite{Farrar18} vector mesons is the most important, because while the scalar meson $f_0(500)$ is lighter, it is a loosely-bound di-pion~\cite{1815806} or tetraquark, hence should have very small coupling to a deeply $S$ which would be very compact, unlike a possible loosely bound di-$\Lambda$ molecule resembling the deuteron and having mass $\approx 2 m_\Lambda$.} Interactions of $S$ can therefore be modeled by the following hadron-level effective Lagrangian 
\begin{equation}
\label{eqn:lagr}
    \mathcal{L}_{\textrm{eff}} \supset \frac{\gt}{\sqrt{40}} \bar{\psi}_B \gamma^5 \psi^c_{B'} \phi_S + g_{SSV}\, \phi_S^\dagger \partial_\mu \phi_S V^\mu + \textrm{h.c.},
\end{equation}
where the superscript $c$ denotes charge conjugation $\psi^c = -i\gamma^2 \psi^*$. The first term is a Yukawa interaction with coupling constant $\gt/\sqrt{40}$. Here, $\gt$ is the dynamical transition amplitude between six quarks in an $S$ and in two baryons as introduced in Sec.~\ref{sec:intro}. The Clebsch-Gordan factor $\sqrt{1/40}$ results from the color-spin-flavor wavefunction projection between $S$ and individual color-singlet di-baryon states~\cite{farrarwintergerst}; details are given in appendix~\ref{app:CSF}.  An important fact is that in the fully-antisymmetrized $S$ wavefunction, only $\sqrt{1/5}$ of the amplitude is contributed by a product of two color-singlet baryons;  the other $\sqrt{4/5}$ is composed of products of color-octet states~\cite{farrarwintergerst}.  That $\sqrt{1/5}$, combined with the fact that there are 8 distinct flavor octet spin-1/2 baryons, accounts for the factor $1/\sqrt{40}$ appearing in eq.~\eqref{eqn:lagr}. The second term in eq.~\eqref{eqn:lagr} accounts for the coupling of $S$ to vector mesons. The phenomenology of the two types of interactions is distinctively different. The Yukawa interaction governs the transition between $S$ and two baryons, and is relevant to processes such as the decay of $S$ to two baryons and formation of $S$ in exclusive scattering reactions. In contrast, the vector vertex gives rise to scattering of $S$ with other hadrons through the exchange of vectors mesons, which can be constrained by DM direct detection experiments and cosmology~\cite{fwx20,farrarxu}. For this paper, we concentrate on the phenomenology arising from Yukawa vertices proportional to $\gt$.

\subsection{Mass and phenomenology}
Theoretical and lattice computations done by different groups yield a wide spread of estimated $S$ mass, ranging from deeply-bound, loosely-bound and unbound~\cite{Jaffe77,Rosner:1985yh,Gignoux:1987cn,Callan:1985hy,Straub:1988mz,Nishikawa:1991di,Oka:1990vx,Haidenbauer:2011za,LQCD11,LQCD13,SasakiHALQCD20,Kodama:1994np}.
On the low end, the recent QCD sum rule calculation~\cite{Azizi19} obtained $m_S\approx 1200$ MeV. Values of $1200$ and $2170$ MeV were found in~\cite{Strumia18} using diquark models with different means of estimating parameters. A prediction of 1750 MeV is made by Ref.~\cite{Evans:2023zde} in holographic QCD. Ref.~\cite{Buccella} obtains a mass $\approx 1890$ MeV by fitting the chromo-magnetic interactions between diquarks.  For other earlier estimates see~\cite{Farrar:2022mih}. The NPLQCD lattice QCD mass calculation~\cite{LQCD13} had a very large pion mass ($\approx 800$ MeV) so is indicative at best, but found binding of about 80 MeV relative to two $\Lambda$'s, consistent with Jaffe's bag model prediction.  
The more recent Mainz group calculation~\cite{greenMainzH21} using the Luscher method reduced the pion mass to below 500 MeV and predicted a weak binding of $\sim {5}$ MeV. 
The HALQCD group finds attraction, but insufficient to form a bound state~\cite{SasakiHALQCD20}.  The HALQCD method seeks to identify bound states by measuring the scattering amplitude of two hadrons and solving the Bethe-Salpeter equation to infer the potential.  It is only applicable to weakly bound states due to their locality approximation~\cite{SasakiHALQCD20}, so would not be expected to detect an $S$.  Moreover the $1/40$ probability of the $S$ appearing in products of color-singlet baryons used -- $\ket{\Lambda \Lambda } $, $\ket{\Sigma \Sigma } $ and $\ket{\Xi N } $ -- discussed above, means that the scattering states explored in all existing lattice calculations have weak overlap with the $S$, making it even harder to see a signal.  

\begin{figure}[tb]
    \centering
    \includegraphics[width=0.6\textwidth]{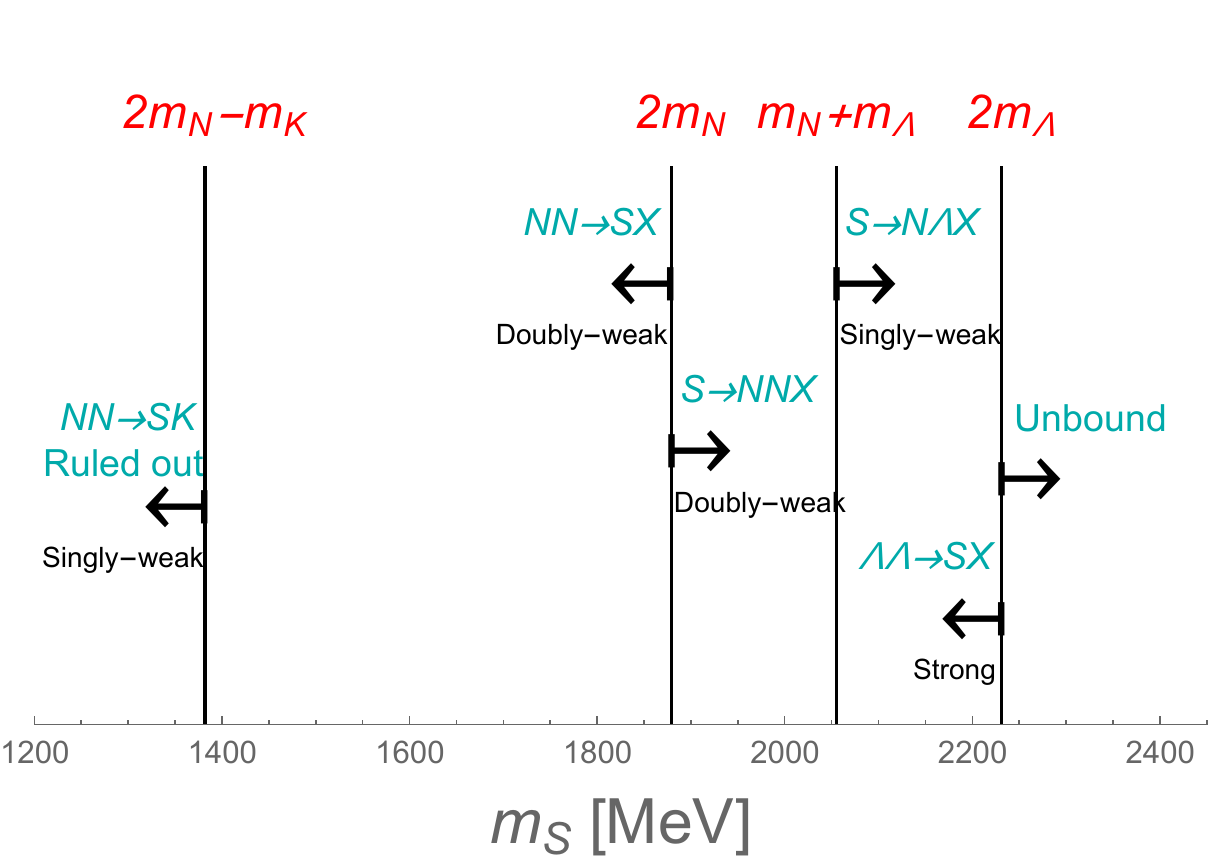}
    \caption{A schematic summary of the landscape of $S$ phenomenology. Each vertical line indicates a critical mass scale where certain reactions are allowed. Note that the mass of a nucleon is coarsely denoted by $m_N\approx 938$ MeV here, with the mass splitting between the proton and neutron, the binding energy of nucleons inside a nucleus, and the mass of additional $e^\pm$ neglected in this figure. From left to right: singly-weak transitions $NN\to SK$, doubly-weak transitions $NN\to SX$; doubly-weak decays $S\to NNX$; singly-weak decays $S\to N\Lambda X$; strong-interaction transitions $\Lambda\Lambda\to SX$; unbound (a broad resonance). }
    \label{fig:landscape}
\end{figure}

The large variance in estimating $m_S$ motivates us to consider the entire possible mass range below $2m_\Lambda$. From lower $m_S$ to higher, the landscape of phenomenology is elaborated below and summarized schematically in figure~\ref{fig:landscape}.  

\begin{enumerate}
    
\item For $m_S<2m_N-m_K\approx 1382$ MeV, two nucleons in a nucleus could be converted to an $S$ and a kaon via a singly-weak process, raising tension with the known stability of nuclei. Besides, stability of neutron stars excludes light baryonic states below 1.2 GeV~\cite{mackeen18,Baym:2018ljz}, and a light scalar $S$ would worsen the instability because it does not have Fermi pressure.\footnote{Ref.~\cite{shahrbaf+22} showed that the observed neutron star mass-radius relationship, and neutron star masses above 2 $M_\odot$, are compatible with the existence of a sexaquark with mass above $2 m_N$ when quark deconfinement at high pressure is taken into account, as indeed is necessary with or without an $S$.} Therefore, we consider $m_S \lesssim 1382$ MeV as observationally ruled out.

\item  If $1382~ \mathrm{MeV} < m_S<2m_N-B_{NN}$, where $B_{NN}$ is the 2-nucleon binding energy, the doubly-weak decay of a nucleus to an $S$ can give observable signatures.  We will use the $e^{\pm}$ searches at the SNO experiment to constrain deuteron decaying to $S$ and thereby place the strongest limits on this mass range. This updates the analysis of~\cite{FarrarZaharijas} based on estimation of the sensitivity of SuperK to ${}^{16}\textrm{O}$ decay.

\item The narrow mass range in which an $S$ is too light to decay to a deuteron or free nucleons, while at the same time is heavy enough that nuclear stability is ensured, $m_D - m_e \leq m_S \leq m_D + m_e$ (or 1875.1 - 1876.1 MeV), is much less constrained. In practice, due to the narrow phase space and detection thresholds of low energy $e^{\pm}$, the mass window 1870 - 1880 MeV is hard to constrain by decay of nuclei or decay of $S$. The cooling of SN1987a and observations of hypernuclei provide the best bounds here.

\item If $m_S\gtrsim 2 m_N$ but $m_S \lesssim m_n + m_\Lambda$, the $S$ can decay to two-nucleon final states via a doubly-weak interaction. There are four principal channels:
\begin{equation}
    \begin{aligned}
    S \to \begin{cases} \deuteron e^-\bar{\nu}_e, \quad &m_{S} \geq 1876.1\, \textrm{MeV}, \\ pp e^- e^- \bar{\nu}_e \bar{\nu}_e, \quad &m_{S} \geq 1877.6\, \textrm{MeV}, \\ np e^- \bar{\nu}_e ,  &m_{S} \geq 1878.3\, \textrm{MeV}, \\ nn ,  &m_{S} \geq 1879.1\, \textrm{MeV}. \end{cases}
    \end{aligned}
    \label{eqn:2N}
\end{equation}
To be a DM candidate, the $S$ lifetime must be at least longer than the age of Universe $t_{\textrm{univ}}=1.37\times 10^{10}$ yr~\cite{Poulin:2016nat}. Additionally, charged leptons in the decay final states may be observed in astrophysical or laboratory measurements, which we use here to set much stronger limits on $\gt$ if $S$ is a significant component of dark matter. If $m_S\gtrsim m_N+m_\Lambda\approx 2055$ MeV, singly-weak decays of $S$ result in a much shorter lifetime.

\item As long as $m_S < 2 m_\Lambda = 2231.4$ MeV, strong-interaction transitions $\Lambda\Lambda \to SX$ are allowed. These transitions can be constrained by observations of double-$\Lambda$ hypernuclei~\cite{hyper1991,hyper2010,hyper20011,hyper20012,FarrarZaharijas} and by their impact on the cooling of supernovae~\cite{McDermott18}.

\end{enumerate}
Generically, the rate of all of the processes above are proportional to $\gt^2$, since they either require the $S$ to disintegrate to or be produced from two baryons. 

\begin{figure}[tb]
     \centering
     \includegraphics[width=0.75\textwidth]{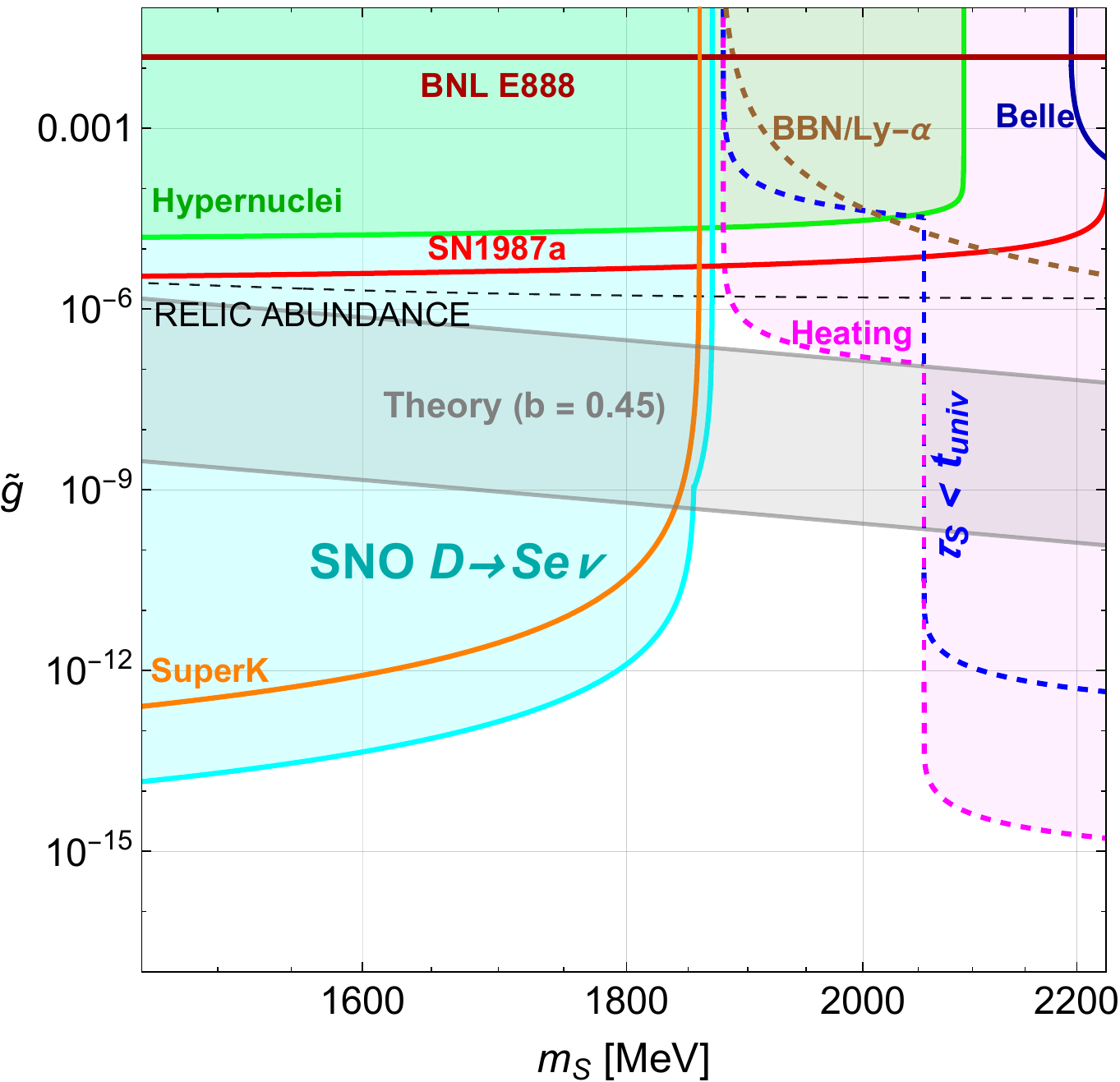}
        \caption{Excluded and predicted regions of $\gt$. The gray band is the theoretical best-estimate for $\gt$.  The solid lines/regions are observational limits on $\gt$ which do not require $S$ to be DM. The dashed lines show DM-based constraints if DM is entirely made of $S$. See main text for detailed description and derivation for each limit. As discussed in the text, the exclusion regions are sensitive to the typical quark momenta in the process, taken to be 100 MeV here.}
        \label{fig:exclusion}
\end{figure}

Observational constraints on $\gt$, which are the main results of our paper, are summarized in figure~\ref{fig:exclusion}. 
Here we briefly describe the figure, postponing the in-depth derivations to sections~\ref{sec:nuclei}--\ref{sec:BBG}.  
The cyan line in figure~\ref{fig:exclusion} follows from our analysis of the deuteron decay $D\to Se\nu$, using data from the SNO experiment (Sec.~\ref{sec:nuclei}). This improves significantly on the estimated upper limit based on SuperK sensitivity to oxygen decay~\cite{FarrarZaharijas}, shown by the orange line. The green region is excluded by the observation of hyperon decay products from doubly-strange hypernuclei, thus placing a limit (reviewed in Sec.~\ref{sec:hypernuclei}) on the formation time of $S$~\cite{FarrarZaharijas}. The red line is adapted from~\cite{McDermott18} and is the upper limit on $\gt$ required for the observed cooling of SN1987a as discussed in Appendix~\ref{app:SN}.  Additionally, in Sec.~\ref{sec:Hdibaryon} we recast the H-dibaryon searches from two accelerator experiments, BNL E888 and Belle; the resulting upper limits on $\gt$ are respectively given by the dark red and dark blue lines.  As noted in the detailed discussions which follow, e.g., Sec.~\ref{sec:deuteron}, some of the bounds depend quite sensitively on the typical momenta of the virtual quarks involved in the process, $p_q$;  in figure~\ref{fig:exclusion} $ p_q$ is fixed to 100 MeV. The gray band is a representative theoretical prediction for $\gt$ based on the Isgur-Karl model of spatial wavefunctions following~\cite{FarrarZaharijas}, reviewed in Sec.~\ref{sec:BBG}.

The bounds on $\gt$ discussed above apply whether or not $S$ is the DM particle.  If $S$ constitutes part or all of the DM, four additional DM-based constraints can be obtained; these are shown by dashed lines in figure~\ref{fig:exclusion}, taking DM to be entirely composed of $S$. The black line is the maximum value of $\gt$ such that $S$ does not come into chemical equilibrium with baryons in the early Universe, so that its relic abundance can be preserved to low redshift~\cite{Farrar18} (see also appendix~\ref{app:relic}). The blue line indicates the value of $\gt$ for which the lifetime of $S$ is equal to the Hubble time. The brown line follows from requiring that the accumulation of deuterium since BBN, due to $S\to De\nu$, be consistent with the measured deuterium abundance in damped Lyman-$\alpha$ systems given the primordial $D$ abundance predicted by LCDM with parameters from the CMB. The magenta region is ruled out by a new limit derived here based on the energy injection of $S$ decay products to astrophysical systems including CMB~\cite{SlatyerWu16}, Lyman-$\alpha$ forests~\cite{Liu:2020wqz}, and gas-rich dwarf galaxies~\cite{WW21}. If $S$ is a fractional component of the DM, the dashed brown and magenta exclusions are weakened by a factor of $\sqrt{f_S}$ where $f_S$ is the fraction of DM composed of $S$; how the dashed black constraint depends on $f_S$ requires a complete study of solving Boltzmann equations and is beyond the scope of this paper. 

\section{Stability of nuclei}
\label{sec:nuclei}

The transition between a two-nucleon state and $S$ (and additional leptons if necessary) must be doubly-weak to convert two units of strangeness. This motivates us to evaluate the matrix elements $\matrixel{S}{H_W}{NN'}$, where $H_W$ is the effective Hamiltonian that characterizes the doubly-weak interaction. To proceed, we can insert a complete di-baryon state
\begin{equation}
    \matrixel{S}{H_W}{NN'} =  \sum_{BB'} \braket{S}{BB'} \matrixel{BB'}{H_W}{NN'}.
\end{equation}
Because $\braket{S}{BB'}$ is only non-zero for eight particular color-singlet di-baryon states, in the exact flavor SU(3) limit where all octet baryons have an equal mass the matrix element sums to zero (see eq.~\eqref{eqn:csf} in appendix). In the presence of flavor breaking, the matrix element is dominated by the lightest di-baryon transition, i.e.,
\begin{equation}
    \matrixel{S}{H_W}{NN'} \approx \frac{\gt}{\sqrt{40}} \matrixel{\Lambda\Lambda}{H_W}{NN'}.
\end{equation}

The non-vanishing matrix elements $\matrixel{S}{H_W}{NN'}$  enable a number of reactions involving $S$ and two nucleons. Notably, if $S$ is light enough, two nucleons in a nucleus could be converted to an $S$ causing nuclei to be unstable. On the other hand, if $S$ is heavy enough, it can decay to two nucleons.

\subsection{Decay of deuterons}
\label{sec:deuteron}
The three leading channels of nuclei decaying to $S$ are
\begin{equation}
\begin{cases}
pp\to See\nu\nu &\Longleftrightarrow \quad  (A, Z) \to (A-2, Z-2) + S ee\nu\nu, \\
np \to Se\nu   &\Longleftrightarrow\quad (A, Z) \to (A-2, Z-1) + S e\nu, \\
nn \to S  &\Longleftrightarrow \quad(A, Z) \to (A-2, Z) + S.
\end{cases}
\end{equation}
The lightest nucleus that could decay to $S$ is the deuteron $D$, via the reaction $\deuteron\to Se\nu$. The ground state of deuterons is dominated by $J=1$ and $L=0$, with a subdominant ($\sim 4$\%) component having $J=1$ and $L=2$~\cite{deuteron}. For our purposes we will neglect this admixture and restrict the discussion to $L=0$.

\begin{figure}[htb]
    \centering
    \begin{subfigure}[b]{0.33\textwidth}
    \centering
    \includegraphics[width=\textwidth]{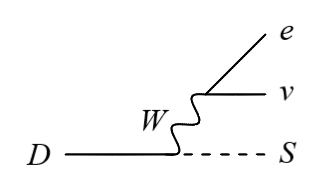}
    \end{subfigure}
    \hfil
    \begin{subfigure}[b]{0.33\textwidth}
    \centering
    \includegraphics[width=\textwidth]{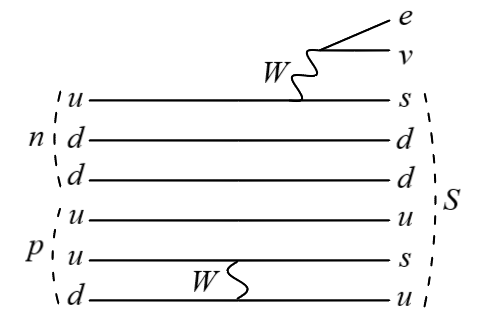}
    \end{subfigure}
    \caption{Feynman diagrams for $\deuteron\to S e^+ \nu$ at the hadron level (left) and the quark level (right).}
    \label{fig:dSenu}
\end{figure}

Since an accurate modeling of $\deuteron\to Se\nu$ at the level of individual nucleons is difficult, we treat the deuteron as an effective vector field $D^\mu$. The decay must be transmitted by two virtual $W$ bosons, with one absorbed internally and one decaying to leptons. Thus, the coupling of $D$ to the $W$ boson and $S$ can be modeled by
\begin{equation}
\label{eqn:dlagr}
    \mathcal{L} \supset \frac{\gt}{\sqrt{40}}\,g_{\deuteron SW}\, \phi_S^\dagger W_\mu \deuteron^\mu + \frac{ig_W}{2\sqrt{2}} W_\mu \bar{\psi}\gamma^\mu (1-\gamma^5) \psi,
\end{equation}
where $g_W$ is the weak coupling constant defined by $G_F=\sqrt{2} g_W^2/(8m_W^2)$. The effective Feynman diagram for this decay at the hadron level is shown in the left panel of figure~\ref{fig:dSenu}. We also show one example of quark-level diagrams in the right panel of figure~\ref{fig:dSenu}.
The coupling constant $g_{DSW}$ encodes the exchange of an internal $W$ boson and emission of the $W$; it has dimension of mass.  We can estimate it by
\begin{equation}
    g_{\deuteron SW} \simeq g_W G_F\, p_q^3\,\sin^2 \theta_C \cos \theta_C ,
\end{equation}
which follows from the bound state Bethe-Salpeter amplitude. Here $\theta_C$ is the Cabbibo mixing angle with $\sin \theta_C = 0.22$, and $p_q$ is the characteristic momenta of quarks. The power of $p_q$ follows from the construction and can be deduced on dimensional grounds based on the mass dimension of $g_{\deuteron SW}$ being one, from the effective Lagrangian eq.~(\ref{eqn:dlagr}). We note that the decay rate $\Gamma \propto g_{\deuteron SW}^2$ is highly sensitive to the numerical value of $p_q$. Throughout the paper, we will proceed with $p_q = \Lambda_{\textrm{QCD}} \simeq 100$ MeV. We note that $p_q$ can take values up to the constituent masses (for light and strange quarks, the constituent mass is roughly 340 MeV and 480 MeV respectively) and thus, the decay rate could be enlarged by a factor of $3.4^6\simeq 1544$. Or conversely, if the mass difference between D and $S$ is small, the effective momenta in the integrals might be smaller than 100 MeV. Therefore, the appearance of $p_q^6$ in the rate raises the major uncertainty in our modeling of $D$ decay, and subsumes other subleading sources of uncertainties, such as the summation of different quark level diagrams. 

We directly compute the decay amplitude from the effective Lagrangian eq.~(\ref{eqn:dlagr}) 
\begin{equation}
    \langle|\mathcal{A}|^2\rangle = \frac{g_{DSW}^2g_W^2}{3m_W^4} (3E_e E_\nu-\vec{p}_e\cdot \vec{p}_\nu),
\end{equation}
and the differential decay rate can be found by
\begin{equation}
    \dv{\Gamma}{E_e d\Omega_e d\Omega_\nu} = \frac{1}{(4\pi)^5 m_D}\int dE_\nu\, \langle|\mathcal{A}|^2\rangle.
\end{equation}
This leads to
\begin{equation}
    \dv{\Gamma}{E_e } = \frac{\gt^2\, G_F^4 p_q^6\, \sin^4\theta_c \cos^2\theta_c}{120\pi^3 m_\deuteron m_S} \sqrt{E_e^2-m_e^2} E_e(m_\deuteron-m_S-E_e)^2.
\end{equation}
Integrating over all possible positron energies $m_e \leq E_e \leq m_\deuteron-m_S$ yields the decay rate
\begin{equation}
\label{eqn:drate}
    \Gamma = \int_{m_e}^{m_\deuteron-m_S}dE_e\, \dv{\Gamma}{E_e }.
\end{equation}

\subsection{Experimental constraints from SNO}
\begin{figure}[htb]
    \centering
    \includegraphics[width=0.7\textwidth]{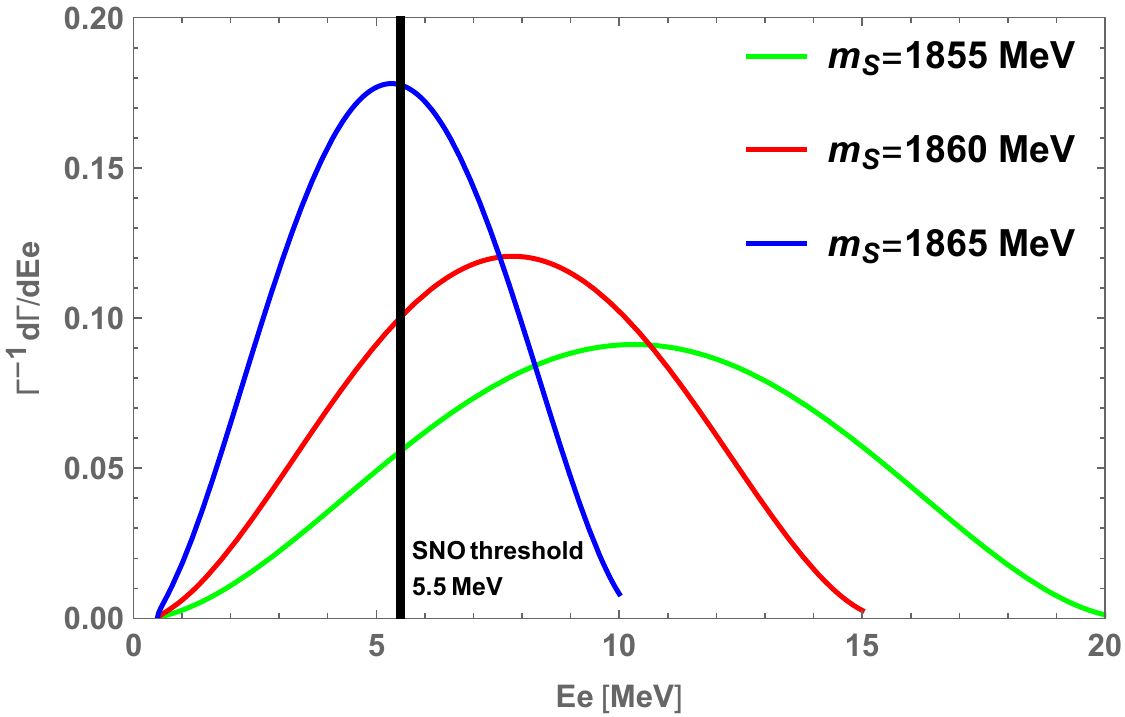}
    \caption{The expected energy spectrum of positrons, $\Gamma^{-1}d\Gamma/dE_e$ in unit of $\textrm{GeV}^{-1}$, produced in the deuteron decay $\deuteron\to Se^+\nu$ for $m_S=1855, 1860$ and $1865$ MeV. The solid black cutoff at $E_e=5.5$ MeV is the threshold of $e^\pm$ detection at SNO. 
    }
    \label{fig:spectrum}
\end{figure}

A stringent laboratory constraint on deuteron instability can be deduced from the SNO experiment. The SNO detector~\cite{SNO} contains $10^6$ kg of heavy water, and the positrons produced in $\deuteron\to Se\nu$ would create Cherenkov radiation which can be detected by photomultipliers (PMTs) installed in the detector\footnote{Although the PMTs are designed to detect electrons produced by solar neutrinos in the reaction $\nu_e+\deuteron\to2p+e^-$, positrons would give the same signal because the spectrum of Cherenkov radiation is only sensitive to charge-squared. 
}.
Figure~\ref{fig:spectrum} shows the expected energy spectrum of positrons from $\deuteron\to Se^+\nu$, for various values of $m_S$. The vertical black line at $E_e=5.5$ MeV indicates the detection threshold. We also define the function $f(x)$ to be the fraction of positrons that have energy above $E_e=x$,
\begin{equation}
    f(x) =  \frac{\int_x^{m_\deuteron-m_S}dE_e\, d\Gamma/dE_e}{\int_{m_e}^{m_\deuteron-m_S}dE_e\, d\Gamma/dE_e}.
\end{equation}
Two important values of $x$ are, \emph{i)} $x=5.5$ MeV is the detection threshold, and \emph{ii)} $x=20$ MeV is the highest energy of $e^\pm$ detected by SNO.

We require the positrons within $5.5\dash 20$ MeV produced by deuteron decay be less than the recorded $e^\pm$ counts at SNO, and in addition, positrons higher than 20 MeV should be statistically consistent with the null result. We also make the conservative assumption that all the observed $e^{\pm}$ events at SNO are due to deuteron decay. Therefore, we can set limits on the decay rate $\Gamma$ by
\begin{equation}
\begin{cases}
   (f(5.5)-f(20))\times N_0(1-e^{-\Gamma t}) < N_{\textrm{obs}}, \\
    f(20)\times N_0(1-e^{-\Gamma t}) < 2.44,
\end{cases}
\label{eqn:snolimit}
\end{equation}
whichever is stronger. Here, $N_0=6.0138\times10^{31}$ is the number of deuterons in the SNO tank, $N_{\textrm{obs}}\approx 2465$ is the observed number of $e^{\pm}$ events recorded during $t=391.432$ days, and $2.44$ in the second equation is the 90\% confidence level upper limit consistent with null observation. With the expressions for $\Gamma$ and $f(x)$ in eqs.~(\ref{eqn:drate},\ref{eqn:snolimit}), this implies lower limits on $\gt$ as a function of $m_S$. The resulting excluded region taking $p_q=100$ MeV is plotted as the cyan area in figure~\ref{fig:exclusion}. We also note that when $m_S\lesssim 1750$ MeV, new decay channels that include pions in the final state open up, so these limits on $\gt$ could be strengthened for $m_S\lesssim 1750$ MeV.  This is not highly motivated because the current limit is so strong relative to theoretical estimates, discussed below, that $m_S\lesssim 1750$ MeV already is strongly disfavored.

\subsection{Other constraints}


Deuteron decay can have other observational consequences in astrophysics. We give two examples below.
\begin{itemize}
    \item The primordial abundance of deuterium, D/H, would be depleted at low-redshift relative to the BBN predicted value. Ref.~\cite{Farrar18} reports that the comparison of predicted and measured values of D/H indicates $\tau_D \gtrsim 5\times 10^{14}$ yr. This is however much less stringent than our SNO analysis which indicates $\tau_D\gtrsim 10^{32}$ yr.
    
    \item Deuterium in the interstellar medium can decay to $Se^+\nu$ and produce positrons with $\order{\textrm{MeV}}$ energies. This provides a potential solution to explain the Galactic 511 keV line observed by the INTEGRAL satellite~\cite{Siegert:2015knp}. Using the INTEGRAL implied upper limit on the positron injection rate $4\times 10^{43}/\textrm{s}$~\cite{DeRocco:2019fjq}, we find $\gt\lesssim 3\times10^{-2}$. At present, this is a very weak limit on $\gt$.
\end{itemize}
Both of these limits are much weaker than the SNO limit and therefore, we refrain from showing them in figure ~\ref{fig:exclusion}.

Heavier nuclei with larger binding energies could also decay in similar ways. For example, Oxygen-16 could decay to Nitrogen-14 by ${}^{16}\textrm{O}\to {}^{14}\textrm{N}+S+X$ if $m_S\lesssim 1862$ MeV. The water tank at SuperK can in principle provide data to constrain the decay. However the existing analyses at SuperK focus on exclusive searches for $e^\pm$ or pions with energy more than $\order{\textrm{100 MeV}}$ (see e.g. \cite{superk,BNV}), and therefore these analyses are not relevant to $S$ for which the positron energy of interest is $\order{\textrm{MeV}}$. Ref.~\cite{FarrarZaharijas} estimated based on the SuperK noise level that an ${}^{16}\textrm{O}$ lifetime less than $\sim 10^{26}$ yr would trigger the SuperK detector. It translates to the orange line in figure~\ref{fig:exclusion} and is superseded by our SNO limits. Future dedicated inclusive searches for di-nucleon decay events could be useful to place improved limits on $S$, and other baryon number violating processes~\cite{BNV} in broader contexts.

\section{Stability of sexaquark dark matter}
\label{sec:S}

Evidence of DM has been observed from both early and late Universe phenomena and therefore, DM is believed to be stable on the cosmological timescale. For this reason, many models of DM make use of symmetries to stablize DM particles so that they are perfectly stable. On the other hand, some popular DM candidate particles can decay and have a finite lifetime, such as axions~\cite{Peccei:1977hh} and sterile neutrinos~\cite{Dodelson:1993je}, as long as the lifetime is consistent with observational constraints. Recently, decaying DM has also been motivated by various tensions in cosmology, including Hubble tension~\cite{Berezhiani:2015yta} and $S_8$ tension~\cite{Enqvist:2015ara,FrancoAbellan:2020xnr}.

\begin{figure}[htb]
\centering
\includegraphics[width=0.8\textwidth]{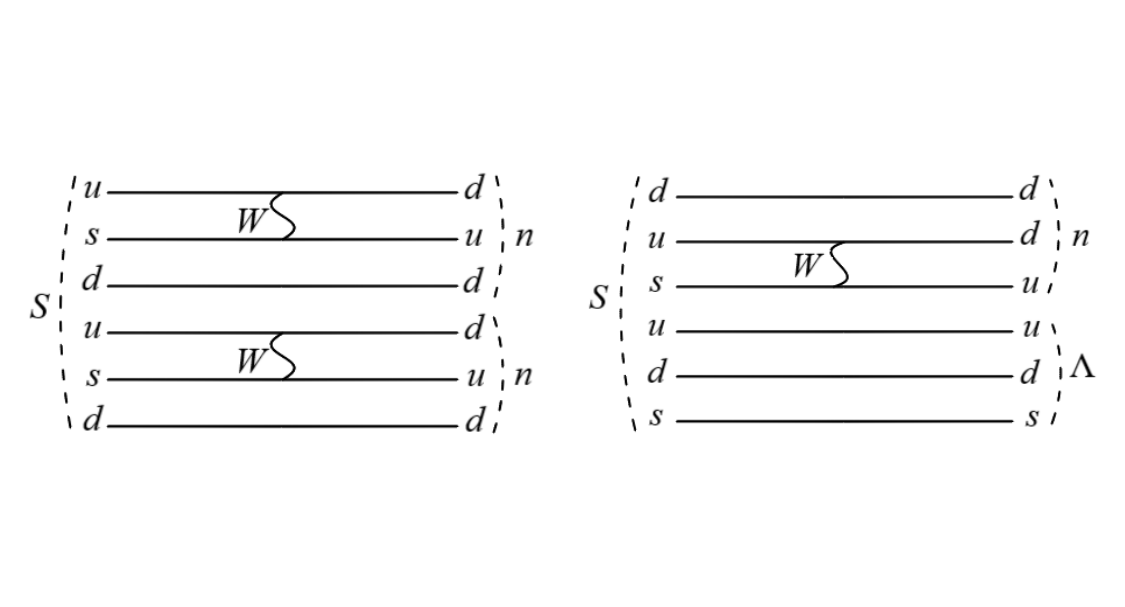}
\caption{Left: An example of a quark diagram for the doubly-weak decay $S\to nn$. Right: An exemplary quark diagram of the singly-weak decay $S\to n\Lambda$.}
\label{fig:Sdecay}
\end{figure}

In this section, we constrain $\gt$ by studying the decay of $S$DM to two baryons. In particular, we will calculate the lifetime of doubly-weak decays $S\to De\nu$ and $S\to nn$, and the singly-weak decay $S\to n\Lambda$. The effective diagram for $S\to De \nu$ is similar to figure~\ref{fig:dSenu}, with the position of $D$ and $S$ swapped. Figure~\ref{fig:Sdecay} shows illustrative Feynman diagrams for $S\to nn$ and $S\to n\Lambda$ at the quark level.

For simplicity, we assume that DM comprises $S$ solely. If $S$ is a fraction $f_S$ of the total DM density, the limits on $\gt$ derived in this section are weakened by a factor of $\sqrt{f_S}$.

\subsection{\texorpdfstring{$S\to nn$}{S to two neutrons}}
\label{sec:nn}

The effective Lagrangian for two-neutron decay of the $S$ is a Yukawa interaction
\begin{equation}
    \mathcal{L} \supset \frac{\gt}{\sqrt{40}} \bar{\psi}_n (g + i g_5 \gamma^5) \psi_n^c \phi_S,
\end{equation}
with the decay rate
\begin{equation}
    \Gamma = \frac{\gt^2 m_{S}}{640\pi} \left[ g^2\left(1-\frac{4m_n^2}{m_{S}^2}\right)^{3/2} + g_5^2 \left(1-\frac{4m_n^2}{m_S^2}\right)^{1/2} \right].
\end{equation}
Since this is a weak decay, one should expect both $g$ and $g_5$ are non-zero. For our purpose here, however, we will take $g_5=0$ which yields the minimal decay rate and therefore the most conservative constraint. We estimate $g$ by the Feynman diagram shown in the left panel of figure~\ref{fig:Sdecay}
\begin{equation}
    g^2 \simeq G_F^4\,p_q^8\, \sin^4\theta_c \cos^4\theta_c,
    \label{eqn:g}
\end{equation}
leading to
\begin{equation}
\label{eqn:snn}
    \Gamma = \frac{\gt^2 G_F^4\,p_q^8\, \sin^4\theta_c \cos^4\theta_c\, m_S}{640\pi} \left(1-\frac{4m_n^2}{m_{S}^2}\right)^{3/2}. 
\end{equation}
As before, we take $p_q$ to be the QCD scale 100 MeV, and therefore $ \Gamma \propto p_q^8$ dominates the uncertainty in our modeling of the decay rate.\footnote{A calculation of weak decays of the H-dibaryon is performed in Ref.~\cite{donoghue86}, implicitly assuming the H is weakly bound.  Thus the momentum dependence can be estimated and $\gt $ is not small. They estimate the weight of different types of quark level interactions that govern the decay.  For us, these would give a sub-leading source of uncertainty relative to $p_q$ and $\gt$.}

There are various astrophysical and cosmological constraints on decaying DM.
\begin{itemize}
    \item Assuming that the lifetime of decaying DM should exceed the age of Universe, $\tau_{\textrm{DM}}>t_{\textrm{univ}}=1.37\times 10^{10}$ yr~\cite{Poulin:2016nat}, from eq.~(\ref{eqn:snn}) we find the blue region in figure~\ref{fig:exclusion} is excluded. However, if $\tau_{\textrm{DM}}=t_{\textrm{univ}}$, about 1/3 DM would have already decayed and the decay products would disturb low redshift astrophysical and cosmological observables. Thus, we can anticipate potentially much stronger limits on the lifetime, as discussed below.
    
    \item Neutrons from $S \to nn$ will undergo $\beta$-decay and produce electrons. The energy of these electrons is peaked around 1 MeV; it is only very weakly sensitive to the value of $m_S$. Constraints on DM decaying to $e^\pm$ pairs have been extensively studied in literature~\cite{essigetal13,SlatyerWu16,Liu:2020wqz,WW21} based on energy injection from $e^\pm$ to astrophysical systems. Among them, the strongest limit for $E_e = 1$ MeV is from the heating of Leo T, requiring $\tau(\mathrm{DM}\to e^+ e^-) > 10^{26}\,$s for $m_{\mathrm{DM}} = 2$ MeV~\cite{WW21}. Recasting this limit for $S$ yields 
    \begin{equation}
        \tau(S\to nn) \gtrsim \left(\frac{\textrm{2 GeV}}{m_S}\right) 10^{23}  \textrm{ s},
    \end{equation} 
    because $S$ is roughly 1000 times heavier and therefore the flux of electrons with $E_e = 1$ MeV is 1000 times smaller. The excluded range of $\gt$ is shown as the pink area in figure~\ref{fig:exclusion}.
    
    
\end{itemize}

\begin{figure}[htb]
    \centering
    \includegraphics[width=0.6\textwidth]{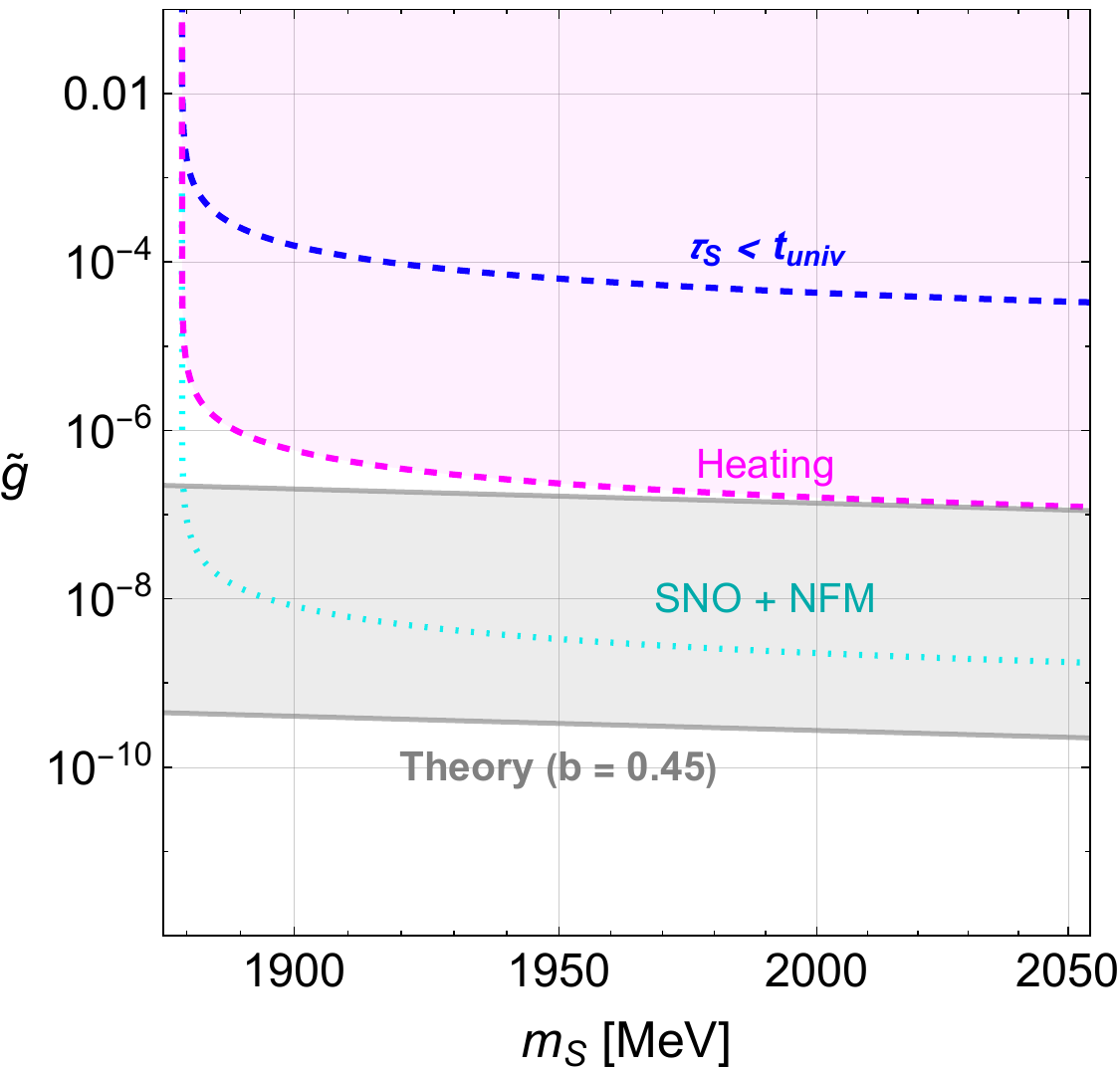}
    \caption{Cyan:  the limit on $\gt$ from the SNO limit on $S\to nn$, taking $p_q = 100$ MeV and assuming DM is composed of $S$, and the ambient density of $S$ is the NFM value $n_S\sim 10^{14}$/cm${}^3$.  This limit is only applicable if the DM-baryon interaction is repulsive, hence it is not included in figure~\ref{fig:exclusion}.  Some other limits from Figure~\ref{fig:exclusion} are shown for comparison.}
    \label{fig:SNONFM}
\end{figure}

A laboratory limit on $S\to nn$ arises if we consider $S$DM with an ambient density $n_S$ in the SNO detector. Phase III of the SNO experiment was equipped with an array of ${}^3$He neutron counters~\cite{SNO} and could observe neutrons produced from $S\to nn$. Assuming all the neutron events at SNO are due to $S$ decay, we require
\begin{equation}
    n_S V \times (1-e^{-t/\tau}) \times 2 \times \epsilon < N_{\textrm{obs}},
    \label{eqn:SnnLim}
\end{equation}
where $V=904.78$ m${}^3$ is the volume of the tank, $\epsilon=0.182$ is the detector efficiency, and $N_{\textrm{obs}}\approx 7000$ is the observed number of neutron events during $t=385.17$ days. To obtain a lower limit on the lifetime $\tau$, we must insert suitable estimations of $n_S$. There are two possibilities:
\begin{itemize}
    \item $n_S\sim 0.1$/cm${}^3$, based on the local DM density of the Galaxy estimated to be $0.2\dash 0.56$ GeV/cm${}^3$ \cite{localdensity} for $m_S\approx 2$ GeV. The corresponding lower limit on the lifetime is $\tau\gtrsim 10^4$ yr, which is far weaker than $t_{\textrm{univ}}$.
    
    \item $n_S\sim 10^{14}$/cm${}^3$, the density of hadronically-interacting DM accumulated in Earth if the effective DM-nucleus cross section is $10^{-30}\dash 10^{-24}$ cm${}^2$~\cite{NFM}, denoted the NFM density hereafter. Such strong DM-nuclei cross sections are in fact not ruled out by direct detection experiments and cosmology due to non-perturbative and finite-size effects~\cite{fwx20,farrarxu}. With this value of $n_S$, eq.~(\ref{eqn:SnnLim}) implies $\tau\gtrsim 4.96 \times 10^{18}$ yr. The corresponding limit on $\gt$ is shown as the dashed cyan line in Figure~\ref{fig:SNONFM}. However the bound does not apply for an attractive $S$-nucleus interaction, since then binding of $S$ with nuclei~\cite{fwx20,farrarxu} kinematically prevents $S$ decay. We therefore do not show this limit in figure~\ref{fig:exclusion}.
    
\end{itemize}

\subsection{\texorpdfstring{$S\to D e \nu$}{S to deuteron}}
\label{sec:lyalpha}
Following the analysis of $D\to Se\nu$ in section~\ref{sec:deuteron}, we can similarly obtain the decay width of $S\to D e\nu$
\begin{equation}
    {\Gamma} =\frac{\gt^2\, G_F^4 p_q^6\, \sin^4\theta_c \cos^2\theta_c}{40\pi^3 m_\deuteron m_S}\int_{m_e}^{m_S-m_\deuteron}dE_e\,  \sqrt{E_e^2-m_e^2} E_e(m_S-m_\deuteron-E_e)^2.
\end{equation}
The decay of $S$DM to deuterons through cosmological history supplies a new source of deuterons in addition to primordial synthesis. Comparing the value of D/H (times $10^{+5}$) measured in damped Lyman-$\alpha$ systems $2.53\pm0.04$~\cite{Cooke:2013cba} and predicted by BBN $2.45\pm0.1$~\cite{Coc:2015bhi}, we conclude $\tau\gtrsim 1.6\times 10^{15}$ yr. The resulting upper limit on $\gt$ is shown as the brown curve in figure~\ref{fig:exclusion}. As a result of the 3-body phase space suppression, it yields a weaker limit on $\gt$ than requiring $\tau(S\to nn)>t_{\textrm{univ}}$, even though the lower limit on lifetime is five orders of magnitude longer than $\tuniv$.

\subsection{\texorpdfstring{$S\to n\Lambda$}{S to neutron+Lambda}}
\label{sec:singlyweak}
If $m_S>2054.47$ MeV, singly-weak decays of $S$ such as $S\to N\Lambda $ are permitted. This is the mass range initially envisaged for the H-dibaryon~\cite{Jaffe77}, where the singly-weak-interaction lifetimes lead to totally different phenomenology and the dibaryon is not a dark matter candidate. For notational convenience, we continue to call the particle $S$.

Among a number of possible decay channels, we only consider $S\to n\Lambda$ because of its simple phase space structure. The process can be described by the effective Lagrangian
\begin{equation}
    \mathcal{L} \supset \frac{\gt}{\sqrt{40}} \bar{\psi}_\Lambda (g+ig_5 \gamma^5) \psi_n^c \phi_S.
\end{equation}
Similar to our discussion of $S\to nn$, we take $g_5=0$ and estimate $g$ by the amplitude of the quark diagram in the right panel of figure~\ref{fig:Sdecay}
\begin{equation}
    g^2 \simeq G_F^2\, p_q^4\, \sin^2\theta_c .
\end{equation}
The decay rate is then given by
\begin{equation}
    \Gamma = \frac{\gt^2G_F^2\, p_q^4\, \sin^2\theta_c}{320\pi} \frac{m_S^2-4m_n m_\Lambda}{m_S} \sqrt{\frac{(m_S^2+m_n^2-m_\Lambda^2)^2}{m_S^4} - \frac{4m_n^2}{m_S^2}}.
\end{equation}
Not surprisingly, much smaller $\gt$ is needed in order for $S$DM to be stable enough once $S\to n\Lambda$ is allowed. The condition $\tau \gtrsim t_{\textrm{univ}}$ can only be met in this mass range for unreasonably low $\gt$, as indicated by the lower part of the blue region in figure~\ref{fig:exclusion}. We also calculate the Leo T heating limit, only considering the heating due to electrons from neutron $\beta$-decay. The result is shown by the lower part of the pink curve in figure~\ref{fig:exclusion}. We note that decay products of $\Lambda$ baryons would also deposit energies to Leo T, making this limit conservative.



\section{Laboratory limits on \texorpdfstring{$S$}{S} production}
\label{sec:Hdibaryon}
The H-dibaryon, a $uuddss$ state with a $\sim80$ MeV binding energy was initially proposed in~\cite{Jaffe77} and lead to many experimental efforts to find it. To date, all experimental searches~\cite{hyper1991,hyper20011,hyper20012,hyper2010,belle,kek1,kek2,BNL1,BNL2} for the H-dibaryon report a null finding. These experiments, as reviewed in~\cite{Farrar17}, either require H-dibaryons to decay to certain final states and therefore must be heavy enough, or the breakup amplitude of H-dibaryons into two $\Lambda$'s (the analogue of $\gt$) is $\order{1}$. These assumptions make the experimental results not address the question of whether a deeply bound configuration of $uuddss$ quarks with a small $\gt$ may exist. Below we review a few representative experiments, and recast their results as upper bounds on $\gt$.



\subsection{Doubly-strange hypernuclei}
\label{sec:hypernuclei}
Hypernuclei that contain two $\Lambda$'s have been created in the laboratory~\cite{hyper1991,hyper20011,hyper20012,hyper2010}. The double-$\Lambda$ hypernucleus is observed to decay weakly to a single-$\Lambda$ hypernucleus which subsequently decays weakly to ordinary nuclei. The lifetime has not been precisely measured, but is assumed to be the typical weak-interaction lifetime $\sim 10^{-10}$ s~\cite{hyper1991}.

The binding energy of two $\Lambda$'s in the nucleus is measured to be $B_{\Lambda\Lambda}\approx 7$ MeV~\cite{hyper20012}. Based on the observed decay pattern of hypernuclei, an H-dibaryon lighter than $2m_\Lambda-B_{\Lambda\Lambda} \approx 2224$ MeV is determined to be excluded because the formation of $\Lambda\Lambda \to H$ with a strong-interaction timescale $\sim 10^{-22}$ s would otherwise occur before the hypernucleus decays. 

\begin{figure}[htb]
    \centering
    \includegraphics[width=0.7\textwidth]{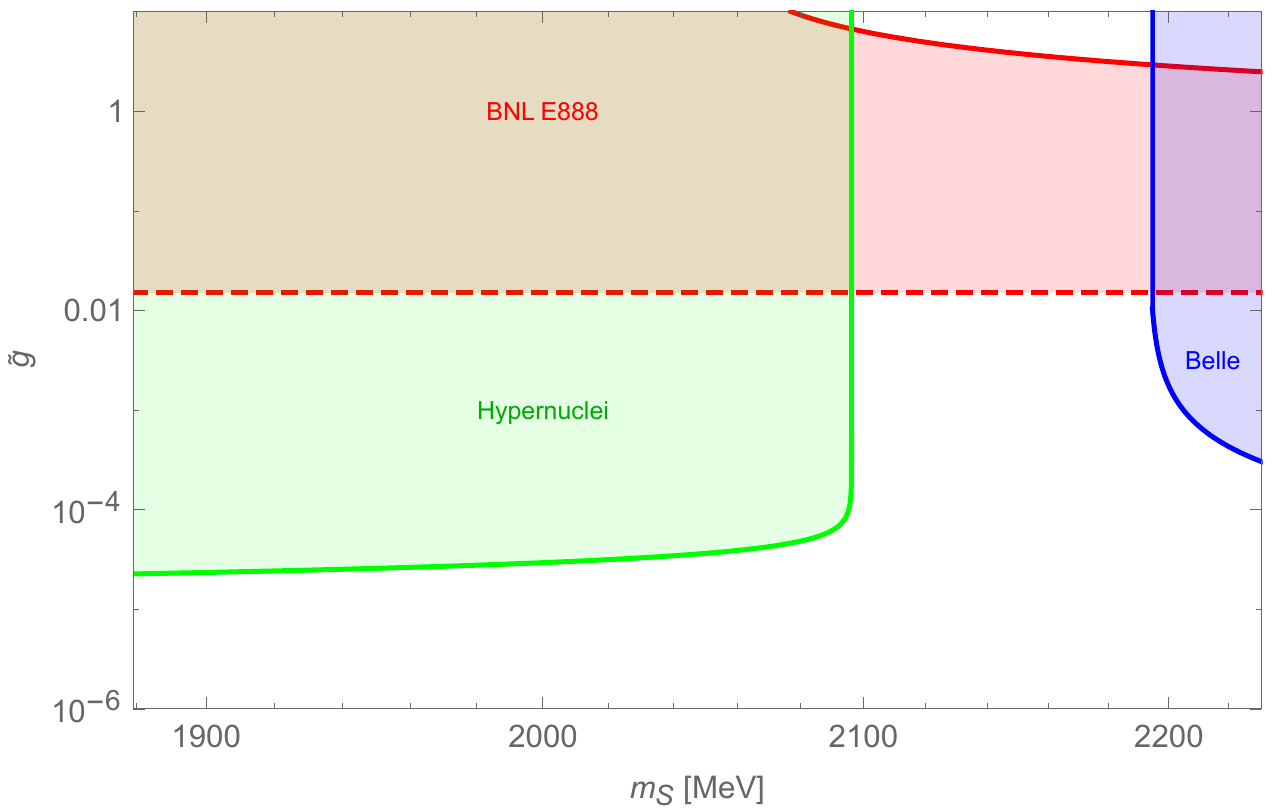}
    \caption{Sensitivity of laboratory H-dibaryon searches to $\gt$.}
    \label{fig:exps}
\end{figure}

These constraints are evaded if the formation rate of $S$ is slower than the weak decay rate of hypernuclei, which is possible if $\gt$ is small enough. The transition timescale of $A_{\Lambda\Lambda}\to A_S$ is worked out in ref.~\cite{FarrarZaharijas}. Requiring it to be longer than $10^{-10}$ s leads to the exclusion region of $\gt$ shown by the green area in Figures~\ref{fig:exps} and~\ref{fig:exclusion}.  The mechanism of $S$ formation $\Lambda\Lambda \to S$ plus additional pions studied in ref.~\cite{FarrarZaharijas} does not apply to masses above 2100 MeV so the excluded region from that analysis ends there, but with further study the hypernuclei limits could be extended to higher mass.

The E836 collaboration at Brookhaven National Lab (BNL)~\cite{BNLE836:1997} searched for evidence of H dibaryon production in the reaction $^3$He($K^-,K^+)Hn$.  The sensitivity was independent of the H lifetime and decay modes, and placed a limit about an order of magnitude lower than the theoretical prediction of~\cite{AertsDoverPRL82,AertsDoverPRD83}.  However since the theoretical calculation assumed the H is quite extended, and moreover ignored the short-distance repulsion of two nucleons, the effective value of $\tilde{g}$ underlying the rate prediction was $\tilde{g} \sim \mathcal{O}(1)$. Thus the limit on  $\tilde{g}$ from E836 is weaker than from the others we discuss. 

\subsection{Diffractive dissociation}
Brookhaven E888 searched for neutral long-lived H-dibaryons~\cite{BNL1}. The experiment assumes H-dibaryons exist in the neutral beam produced by 24.1 GeV proton-Pt collisions, and a scintillator is placed at a distance of $10^{-8}$ s downstream of the Pt target. The reaction of interest is the diffractive dissociation of an H-dibaryon into two $\Lambda$'s by the scintillator, which are detected by the $\Lambda$ decay product to $p\pi^-$
\begin{equation}
    H+A \to \Lambda\Lambda A \to p\pi^- p\pi^- A.
\end{equation}
Similar signals can be triggered by neutrons
\begin{equation}
\begin{aligned}
    n+A & \to \Lambda\Lambda X \to p\pi^- p\pi^- X, \\
    n+A & \to \Lambda K_S X \to p\pi^- \pi^+\pi^- X.
\end{aligned}
\end{equation}
A 90\% confidence level upper bound on the product of $\sigma_S$ and $\sigma_{\Lambda\Lambda}$, the diffractive dissociation cross section of the $H$, is placed at~\cite{BNL1, Farrar17}
\begin{equation}
\label{eqn:diffrac}
    \left.\dv{\sigma_H}{\Omega}\right|_{65\, \textrm{mr}} \frac{\sigma_{\Lambda\Lambda}}{0.5\, \mathrm{mb}} < 2.3\times 10^{-4} \left.\dv{\sigma_n}{\Omega}\right|_{65\, \textrm{mr}} \frac{\sigma_{\Lambda K_S}}{5.9\, \mu \mathrm{b}} ,
\end{equation}
where $\sigma_n$ is the production cross section of neutrons, and $\sigma_{\Lambda\Lambda}$ and $\sigma_{\Lambda K_S}$ are the corresponding diffractive dissociation cross sections of $H$ and $n$ respectively.
Translating this bound to an exact upper limit on $\gt$ requires a detailed calculation of $\sigma_S$ and $\sigma_{\Lambda\Lambda}$ as a function of $\gt$. At this point, we only employ a naive estimation $\sigma_S \approx \sigma_n \gt^2$ so eq.~(\ref{eqn:diffrac}) implies $\gt\lesssim 1.5\times 10^{-2}$, shown by the dashed red line in figure~\ref{fig:exps}. A caveat is that, if $\gt$ is large enough so that the lifetime of $S$ is shorter than $10^{-8}$ s, the $S$ would decay before entering the scintillator. This boundary is shown by the solid red line in figure~\ref{fig:exps}.

\subsection{Upsilon decay}
The $\Upsilon$ decays dominantly into three gluons, and can subsequently source $qqqqqq+\bar{q}\bar{q}\bar{q}\bar{q}\bar{q}\bar{q}$ final states. It is possible that the $6q$ (or $6\bar{q}$) state forms an $S$ (or $\bar{S}$) if their quantum numbers are identical to the $S$ ($\bar{S}$). The formation rate of $S$ from $\Upsilon$ decay is thus independent of $\gt$, and electron-positron colliders such as BaBar and Belle can provide discovery opportunities. However, the probability that the six quarks are singlet under color-spin-flavor is statistically penalized. Ref.~\cite{Farrar17} estimates that the inclusive branching ratio to be $BR(\Upsilon\to SX)\approx 2.7\times 10^{-7}$. 


An inclusive search for the H-dibaryon from $\Upsilon$ decay was performed by Belle~\cite{belle}. The signal being sought was the decay of H into $\Lambda p \pi^-$. An upper bound on
\begin{equation}
    BR(\Upsilon\to HX) \times BR(H\to \Lambda p \pi) \lesssim 10^{-7}
    \label{eqn:belleH}
\end{equation}
was obtained, applicable if $m_S>m_p+m_\Lambda+m_\pi=2193$ MeV.  

To recast the Belle limit for $S$, we must take into account that the $S$ could be much longer-lived than H due to a small $\gt$. This leads to the possibility that $S$ could escape the detector before decaying to $\Lambda p \pi$. Thus, we re-interpret eq.~\eqref{eqn:belleH} by
\begin{equation}
    BR(\Upsilon\to SX) \times BR(S\to \Lambda p \pi) \times \frac{L}{l(S\to \Lambda p \pi)} \lesssim 10^{-7},
    \label{eqn:belleS}
\end{equation}
where $L\approx 3$ m is the detector size and $l(S\to \Lambda p \pi) = \beta \gamma c \times \tau(S\to \Lambda p \pi)$ is the decay length. To proceed, we take $BR(\Upsilon\to SX)= 2.7\times 10^{-7}$, and estimate $BR(S\to \Lambda p \pi)$ by the 3-body phase space suppression factor $\Gamma(S\to \Lambda p \pi)/\Gamma(S\to n\Lambda) \sim 1/(8\pi^2)$. The lifetime of $S\to \Lambda p\pi$ is a function of $m_S$ and $\gt$; details are given in appendix~\ref{app:belle}. The blue region in figure~\ref{fig:exps} shows the exclusion range of $\gt$ from the Belle experiment, applicable for $m_S>m_p+m_\Lambda+m_\pi=2193$ MeV.  Unfortunately, this mass range is not of interest for $S$DM.

\subsection{Cooling of SN1987a}
In addition to the laboratory constraints on $\gt$ discussed above and in figure~\ref{fig:exclusion}, a limit from the cooling time of SN1987a was obtained by McDermott et al~\cite{McDermott18}.\footnote{See however~\cite{Bar19}, where the $\order{10}$ s cooling time of SN1987a is challenged. The SN1987a constraint on $S$ may therefore need to be revised.} We adapt their calculation and find the maximum allowed value of $\gt$ such that the cooling time of SN1987a is not shorter than $\order{10}$ s. This is shown by the red line in figure~\ref{fig:exclusion}. Details are given in Appendix~\ref{app:SN}. 

\section{Isgur-Karl model of \texorpdfstring{$\gt$}{g}}
\label{sec:BBG}

In this section, we derive the gray band in figure~\ref{fig:exclusion} which represents the theoretical modeling of $\gt$, following the methodology of Ref.~\cite{FarrarZaharijas} (\citetalias{FarrarZaharijas} below) and~\cite{fwx20}. Effectively, $\gt$ is the Yukawa coupling of the flavor-conserving $SBB'$ vertex which naively could be expected to be $\order{1}$ since the interaction is hadronic.  However \FZ\ argued that this naive intuition is incorrect because $\gt$ is the transition amplitude between the states: $\gt \equiv |\matrixel{S}{H_{\textrm{QCD}}}{BB'}|$. \FZ\ pointed out that $\gt$ is suppressed on account of several factors:  i) the well-known strong short-distance repulsion between two baryons due to Fermi statistics, ii) the relatively small spatial size of the $S$, and iii) the necessity that all 6 quarks fluctuate between rather orthogonal initial and final states.  Below we first summarize the \FZ\ calculation, then point out an additional effect due to the tunneling-amplitude (analogous to the Gamov suppression factor in nucleosynthesis) which was not included by \FZ.

In the leading approximation of ignoring interactions, setting $H_{\textrm{QCD}} \rightarrow 1$, $\gt$ is just the overlap of the spatial wavefunctions of $S$ and $BB'$.  The overlap was evaluated in \FZ, taking the wavefunctions of the quarks in the baryons and $S$ to be solutions of harmonic oscillator potentials, which are entirely specified by their root mean square radii, $r_B$ and $r_S$.  This is a simple extension of the Isgur-Karl model of baryons~\cite{Isgur:1978wd}. The overlap also depends on the relative wavefunction of the baryons, $u(a)$, where $a$ is the separation between the centers of mass of the two baryons. The general expression for the overlap is given in \FZ 
\begin{equation}
\label{eqn:BBG}
    |\mathcal{M}| = \left(\frac{3}{2\pi r_S^2}\right)^{3/4}\left(\frac{2r_S/r_B}{1+(r_S/r_B)^2}\right)^6\times \int_{0}^{\infty} d^3a\, \frac{u(a)}{a} \exp(-\frac{3}{4}\frac{a^2}{r_S^2}).
\end{equation}
 
A practical choice for $u(a)$ is the Brueckner-Bethe-Goldstone (BBG) wavefunction, which describes the state of two asymptotically free particles in a repulsive short distance potential having a hardcore radius $r_c$.\footnote{To describe the transition within a nucleus, as for the hypernuclear experiments, one should use the relative wavefunction of two nucleons in a nucleus.  This is discussed in \FZ, where it is found that the transition amplitude is only weakly sensitive to the size of the nucleus for larger nuclei. This justifies our approximating the transition amplitude in all cases by the single parameter $\gt$, in our phenomenological discussions above.} 
In the BBG approximation, the baryons are free except for an infinite potential barrier between them below a hardcore radius $r_c$.  Thus $r_c$ sets the lower bound on the separation of the baryons' centers of mass, $a$.  The short distance behavior of the inter-baryon potential is crucial for determining $\gt$ and has often been incorrectly modeled in the literature on $\gt$, so we devote section \ref{sec:criticsRc} to a review of present knowledge of $r_c$.  For our calculations we use the BBG wavefunction with the conservative choice $r_c = 0.4$ fm. 



In order to evaluate Eq.~\eqref{eqn:BBG} we require an estimate of $r_S$. Due to the non-coupling of $S$ to pions and other flavor-octet pseudoscalar mesons (except through an off-diagonal transition to heavier flavor-octet di-baryons),  $r_S$ can be expected to be considerably smaller than $r_p$. An empirical radius-mass relationship based on the Compton wavelengths of a hadron and the lightest meson it couples to is~\cite{Farrar18}
\begin{equation}
\label{eqn:rS}
    r(m) = \frac{1}{m} + \frac{b}{m'},
\end{equation}
where $b$ is a non-negative coefficient, and $m'$ is the mass of the lightest meson that strongly couples to the baryon. Fitting the charge radius of protons $r_p=0.87$ fm with $m'=m_\pi=135$ MeV gives $b=0.45$. 
The second term in eq.~\eqref{eqn:rS} accounts for the fact that the pion cloud makes a significant contribution to $r_p$. 
The relationship eq.~\eqref{eqn:rS} may be generalized to other hadrons with $b$ varying from 0 to 0.45, depending on its coupling to mesons. The $S$ is expected to couple most strongly to the flavor-singlet combination of $\omega$ and $\phi$ vector mesons~\cite{Farrar18} with $m' \approx 1000$ MeV, giving $r_S \lesssim 0.2$ fm.\footnote{In~\cite{farrarwang23}, we use an independent method based on QCD sum rules which agrees with $r_S \ll r_p$.}
Using eq.~(\ref{eqn:rS}), we can then recast the expression eq.~\eqref{eqn:BBG} into a function $\mathcal{M}(m_S)$; for our estimates we conservatively adopt $b=0.45$. 

In addition to the wavefunction overlap, an additional factor may suppress $\gt$, namely the tunneling of the 6-quark configuration between $BB$ and $S$. Overall, $\gt$ is given by
\begin{equation}
\label{eqn:gt}
    \gt = e^{-\mathcal{S}} \times |\mathcal{M}|.
\end{equation}
The tunneling action $\mathcal{S}$ can be estimated as follows~\cite{fwx20}. The characteristic momentum $p_q$ of each constituent quark varies from the QCD scale, $\sim$100 MeV, to the constituent quark mass scale ($\gtrsim$ 340 MeV); the timescale $\Delta t$ for the transition is roughly the typical hadronic length-scale ($\order{1}$ fm) divided by the speed of light. Summing the action $\mathcal{S}\approx p_q \Delta t$ over six quarks gives $e^{-\mathcal{S}}=10^{-4}\dash 0.05$. The gray band in figure~\ref{fig:exclusion} shows the range of $\gt$, corresponding to $r_c=0.4$ fm and $b=0.45$, with the width of the band set by the range of estimated tunneling suppression factor $e^{-\mathcal{S}}=10^{-4}\dash 0.05$. Changing $b$ to a smaller value would decrease $r_S$ and further suppress $\gt$. For a conservative choice, we take $b=0.45$ and only show $\gt$ with $b=0.45$ in figure~\ref{fig:exclusion}.

We stress that a more detailed estimate of the tunneling suppression of $\gt$ is warranted.  For example, if the $S$ is made of three scalar di-quarks $S = \epsilon_{\alpha\beta\gamma} (ud)^\alpha(us)^\beta(ds)^\gamma$~\cite{Strumia18,Buccella}, the transition $S\to BB$ must unbind at least two of the three di-quarks to rearrange the system into two color-singlet baryons. This would imply a more highly suppressed tunneling action if the di-quark binding is as strong as some estimates indicate~\cite{Strumia18,Buccella}.

\section{Critics review}
\label{sec:critics}
In this section, we review existing literature where negative arguments against $S$ as a stable hadron or $S$ as DM were presented. We also compile from existing literature the potential solutions to those negative arguments. As these arguments are not completely conclusive at this stage, discovering or excluding $S$ is still an open and continuing quest. 

\subsection{Short-distance repulsion between baryons}
\label{sec:criticsRc}
A recurring source of divergence in conclusions about present constraints on the $S$ arises from treatment of the short distance potential between baryons, which strongly influences the wavefunction overlap calculation of $\gt$ discussed in the previous section.  The existence of a short distance repulsion between baryons was one of the first properties of hadronic interactions to be observed.  It is responsible for the fact that nuclear radii scale as $A^{1/3}$ to good accuracy, and for the approximate incompressibility of nuclear matter.  We now understand that this so called hard-core repulsion is due to Fermi statistics at the quark level.  In the widely-used Brueckner-Bethe-Goldstone (BBG) wavefunction, the nucleons are taken to be free when their centers of mass are separated by more than the hard-core radius $r_c$, within which the potential is infinite.  The value of $r_c$ is constrained by nucleon-nucleon phase shifts; the fits indicate $r_c\approx 0.4$ fm, with $r_c<\,$0.3 fm or $r_c>\,$0.5 fm being experimentally disfavored~\cite{hardcore}.  Many other well-known NN potentials also include a hard core; for a recent review see~\cite{Naghdi:2007ek}. These potentials are constructed based on measurements of phase shift and NN scattering. For example, the Hamada-Johnston potential~\cite{HAMADA1962382} has a hard core radius 0.343 fm and the Reid potential~\cite{Reid1968} has a hard core radius near 0.4 fm. The potential between two baryons can also be found using lattice QCD.  Figure~1 of~\cite{SasakiHALQCD20} shows that below $a = 0.6$ fm (where $a$ is the separation) the $\Lambda \Lambda$ potential is repulsive, with the repulsion increasing very rapidly at smaller separation to exceed 125 MeV below $a = 0.4$ fm where the plot cuts off.  Thus the hard-core BBG approximation with no repulsion at distances larger than $r_c = 0.4$ fm gives a conservative estimate of the suppression in $\gt$ due to the wavefunction overlap.  Calculations of the wavefunction overlap not taking into account the hard-core repulsion, such as in~\cite{Strumia18, McDermott18} which computed $\gt$ based on the Isgur-Karl model following~\cite{FarrarZaharijas} but ignored hardcore repulsion,  grossly overestimate $\gt$. 

\subsection{$S$ and oxygen stability}
Ref.~\cite{Strumia18} computed $\gt$ following the methodology of~\cite{FarrarZaharijas} (described in Sec.~\ref{sec:BBG}) but used a nucleon-nucleon (NN) wavefunction interpolated from~\cite{Lonardoni:2017egu}.  However the focus of Ref.~\cite{Lonardoni:2017egu} is the form of the wavefunction at large distances and their wavefunction was not constrained to have the known hard-core at short distances.  The authors of~\cite{Strumia18}, by comparing their value of $\gt$ calculated without taking the hard core into account, to the estimated limits on $\gt$ from the oxygen stability in the SuperK detector given in~\cite{FarrarZaharijas}, concluded that $S$ is ruled out for $m_S \lesssim 1860$ MeV, the mass below which two nucleons in an oxygen atom can convert to an $S$.  It is noteworthy that Ref.~\cite{Strumia18} also calculated $\gt$ using the BBG wavefunction with a hardcore radius 0.5 fm and found $\gt$ can be suppressed down to $10^{-10}$ agreeing with~\cite{FarrarZaharijas}. 

The determination of $\gt$ from spatial wavefunctions is extremely sensitive to whether the hard core is included in the NN wavefunction as discussed in Sec.~\ref{sec:criticsRc}. Studies which do not take the hard core into account therefore obtain inaccurate (overly strong) constraints on the allowed properties of the $S$.

\subsection{$S$ and neutron stars}
The impact of $S$ on the cooling of SN1987a is studied in~\cite{McDermott18}, where it was required that in the hot and dense baryonic environment of the proto-neutron star, the timescale of the reaction $\Lambda \Lambda \leftrightarrow S \gamma$ must be longer than the observed cooling time $\order{10}$ s. (In appendix~\ref{app:SN}, we recast their calculation to obtain the supernova cooling upper limits on $\gt$.) The authors of~\cite{McDermott18} then presented two arguments against the existence of $S$. (i) The authors computed $\gt$ following~\cite{FarrarZaharijas} in the Isgur-Karl model using a Brueckner-Bethe-Goldstone NN wavefunction without a hard core, and found that the estimated value of $\gt$ is too large to be compatible with an $\order{10}$ s cooling time. (ii) It was also qualitatively argued that the presence of $S$ inside neutron stars would soften the equation of state, and thus be severely challenged by the existence of neutron stars with masses above two solar masses. Argument (i) is obviated by including a hard core in the NN wavefunction, as discussed in Sec.~\ref{sec:criticsRc}. Argument (ii) was addressed in ref.~\cite{Shahrbaf:2022upc}, which showed that quark de-confinement in the center of massive neutron stars gives good agreement with neutron star observations and is favored phenomenologically with or without the existence of a deeply bound $S$.

\subsection{Relic abundance of $S$}
As discussed in Sec.~\ref{sec:intro}, Ref.~\cite{KolbTurner18} studied the thermal evolution of $S$ and reported that the relic density of $S$ today is too small to make up a significant portion of the DM, unless (i) $m_S \lesssim 1.2 $ GeV or (ii) the breakup cross section of $XS \to BB'$ is much smaller than QCD-scale cross sections. The possibility (i) is not feasible as an $S$ lighter than 1.2 GeV would destablize nuclei by converting nucleons to $S$ and kaons. The possibility (ii) was studied in~\cite{Farrar18}, where the condition $\gt \lesssim 2\times 10^{-6}$ was derived in order for $S$ not to thermalize in the early Universe. This is detailed in appendix~\ref{app:relic}.

\subsection{Direct detection, CMB and $S$ DM}
The direct detection, astrophysical and CMB signals of $S$ DM (assuming $S$ is the the entirety of DM) were studied in depth in~\cite{farrarxu}. It was shown that due to a combination of overburden, non-perturbative phenomenon and extended size of nuclei, for GeV-scale DM a DM-nucleon cross section $\lesssim 10^{-26}$ cm${}^2$ is not excluded by current limits.  A general difficulty for direct detection of  DM in the $\mathcal{O}$(GeV) mass range, is that DM with typical virial velocity $v\sim 10^{-3}c$ often does not carry enough energy to pass trigger threshold, even if it does scatter in the detector.  An ingenious observation is that there should be a population of DM with up to $\approx$ GeV kinetic energy, as a result of upscattering by cosmic rays~\cite{BringmannPospelovCRDM19}.  The recent study~\cite{Alvey:2022pad} places limits on "CRDM" lying in the sweet spot of having small enough cross-section to reach the Xe1T detector in the Gran Sasso National Lab without too much energy loss in the overburden.  Using a perturbative approximation, ~\cite{Alvey:2022pad} concluded that the $S$ could be excluded.  However for the coupling and mass range of interest for the $S$, non-perturbative effects are strong and lead to a short energy-loss length, such that $S$CRDM do not in general reach Xe1T with sufficient energy to be detected; for details see ~\cite{fx23}. 

\section{Summary}

To conclude, we have derived observational constraints on $\gt$, the sexaquark dissociation amplitude into two baryons, as a function of its mass.  Our new, most constraining limit on $\gt$ comes from SNO limits on $e^+$ appearance, applicable if $S$ is lighter than deuterium by more than 5.5 MeV, the SNO threshold for positron detection.  This significantly improves the previous best limit obtained in~\cite{FarrarZaharijas} based on oxygen decay in the SuperK detector.  Comparing to theoretical estimates for $\gt$, we conclude that $m_S < 1.80$ GeV is strongly disfavored based on the stability of deuterium. For $m_S \gtrsim 2m_N \approx 1.88$ GeV, the strongest limits on $\gt$ are from supernovae and hypernuclei experiments, with the latter requiring $\gt \lesssim 10^{-5} \mathrm{-} 10^{-6}$ depending on the value of $m_S$.  We also reviewed the existing laboratory searches for the H-dibaryon and concluded that the best limits exclude $\gt \gtrsim 10^{-5}$. We set additional limits on $\gt$ that are applicable if the $S$ makes up all, or in some cases just part of, the dark matter. In particular, we obtain new, much stronger limits for $m_S \gtrsim 2m_N \approx 1.88$ GeV, required to avoid the decay of $S$ to baryons overheating astrophysical environments.  These are the most constraining limits for $m_S\gtrsim 1.9$ GeV, unless $S$'s constitute $\approx 10$\% or less of the dark matter.

\begin{acknowledgments}

We thank Hartmut Wittig for comments on lattice QCD calculations. ZW thanks Aksel Hallin and Tony Noble for discussions on the sensitivity of the SNO detector to positrons, and Xuyao Hu, Di Liu, Sam McDermott, Marco Muzio, Po-Jen Wang and Xingchen Xu for helpful conversations. Feynman diagrams in this paper are generated by a Feynman diagram maker developed by Aidan Randle-Conde.  The research of GRF was supported by NSF-PHY-2013199 and the Simons Foundation. The work of ZW was suppoted in part by NSF-PHY-2013199, MacCracken Fellowship and James Arthur Graduate Fellowship.

\end{acknowledgments}

\appendix
\section{Non-thermalization of \texorpdfstring{$S$}{S} in the early Universe}
\label{app:relic}
In this appendix we calculate the range of $\gt$ so that $S$ does not chemically thermalize with baryons in the hot hadronic early Universe. We consider processes that convert $S$ to two-baryon states, $K^+ S\to p\Lambda$, $\pi^\pm S\to \Sigma^\pm \Lambda$, $\gamma S\to \Lambda\Lambda$ and $\pi\pi S \to \Lambda\Lambda$. Note that $\pi S \to \Lambda\Lambda$ is forbidden by isospin conservation. The amplitudes of the first two reactions can be found in~\cite{Farrar18}, that of the third in~\cite{McDermott18}, and we scale the rate of $\pi\pi S \to \Lambda\Lambda$ given in~\cite{KolbTurner18} by $\gt^2/40$. We then follow the procedure in~\cite{cannoni} to calculate thermal cross sections. 
\begin{figure}[htb]
    \centering
    \includegraphics[width=0.6\textwidth]{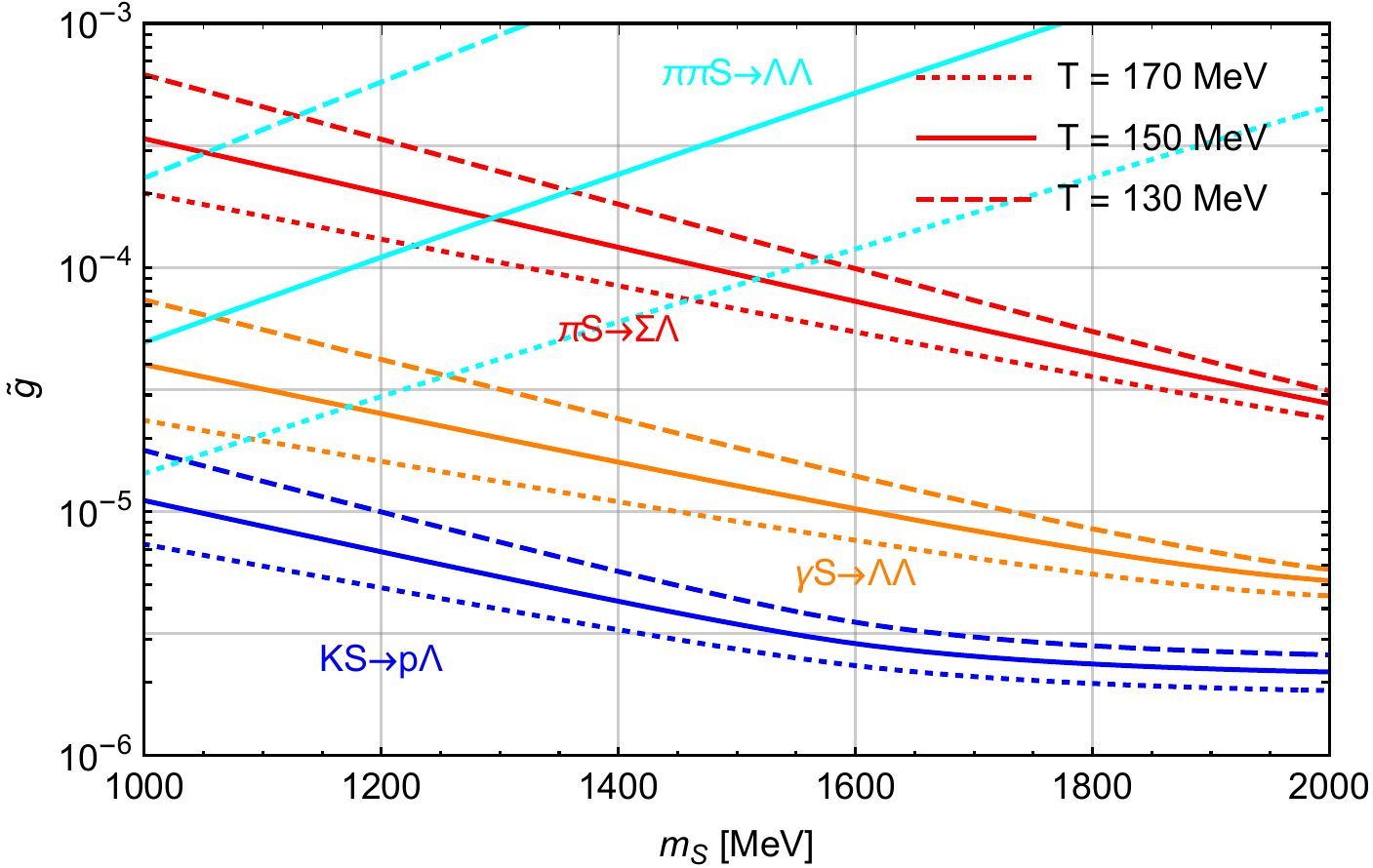}
    \caption{Maximum values of $\gt$ for which $n_X \expval{\sigma v}_{XS\to BB}<H$. Dashed, solid and dotted lines assume $T=130$, $150$ and $170$ MeV.}
    \label{fig:relic}
\end{figure}
Figure~\ref{fig:relic} shows the maximum values of $\gt$ for which $n_X \expval{\sigma v}_{XS\to BB}<H$. In our calculation, we have assumed that the transition from the quark-gluon plasma phase to the hadronic phase takes place abruptly at a temperature $T$, which we vary from 130 to 170 MeV. 
We also find that the dominant breakup process is $KS\to p\Lambda$, despite $n_K$ being smaller than $n_\pi$ and $n_\gamma$. The suppression of $\pi S\to \Sigma\Lambda$ arises from a flavor SU(3) cancellation~\cite{Farrar18}, and $\gamma S\to \Lambda\Lambda$ from the magnetic moment of $\Lambda$~\cite{McDermott18}. For $\gt \lesssim 2\times 10^{-6} $, all of these reactions are slower than the Hubble rate. 

\section{Color-spin-flavor wavefunction projection}
\label{app:CSF}
The normalization factor $\gt/\sqrt{40}$ in eq.~(\ref{eqn:lagr}) is chosen as follows. The total transition amplitude between an $S$ and a two-baryon state, $\braket{S}{BB'}$, consists of two contributions. Firstly, we denote the dynamical amplitude of six quarks transitioning between $S$ and two baryons by
\begin{equation}
    \gt = \matrixel{S}{H_{\textrm{QCD}}}{BB'}.
\end{equation}
Secondly, the projection of color-spin-flavor (CSF) wavefunctions between $\ket{S}$ and $\ket{BB'}$ involves Clebsch-Gordan decomposition. The $S$ is a singlet under color $\textrm{SU(3)}$, spin $ \textrm{SU(2)}$ and flavor $\textrm{SU(3)}$ separately, with its CSF wavefunction given by~\cite{farrarwintergerst}
\begin{equation}
    \ket{S} = \frac{1}{16\sqrt{3}}(\epsilon_{\alpha\mu\rho}\epsilon_{\beta\nu\sigma}\epsilon_{im}\epsilon_{jk}\epsilon_{ln}-\epsilon_{\alpha\beta\rho}\epsilon_{\mu\nu\sigma}\epsilon_{im}\epsilon_{jl}\epsilon_{kn}) u^\dagger_{\alpha,i}u^\dagger_{\beta,j}d^\dagger_{\mu,k}d^\dagger_{\nu,l}s^\dagger_{\rho,m}s^\dagger_{\sigma,n}\ket{\Omega},
\end{equation}
where Greek (Latin) letters indicate color (spin) indices, $u^\dagger,d^\dagger,s^\dagger$ are the field operators for up, down and strange quarks, and $\ket{\Omega}$ is the QCD vacuum. Similar expressions for baryons can be found in e.g.~\cite{thomsom}. Equivalently, the representations of $S$ and baryons under CSF can be depicted by the Young diagrams listed in table~\ref{tab:young}. The eight di-baryon configurations containing a flavor-singlet component are $\Lambda\Lambda, \Sigma^0\Sigma^0, \Sigma^+\Sigma^-, \Sigma^-\Sigma^+, p\Xi^-, \Xi^- p, n\Xi^0$ and $\Xi^0 n$~\cite{McNAMEE}. Projecting the CSF wavefunctions of di-baryons to $S$ leads to a factor of $\pm1/\sqrt{40}$. Thus, the total transition amplitudes are~\cite{farrarwintergerst}
\begin{equation}
\label{eqn:csf}
\begin{aligned}
    & \braket{S}{\Lambda\Lambda} = \braket{S}{\Sigma^0\Sigma^0} = -\braket{S}{\Sigma^+\Sigma^-} = -\braket{S}{\Sigma^-\Sigma^+} = \braket{S}{p\Xi^-} \\
      = & \braket{S}{\Xi^- p} =- \braket{S}{n\Xi^0} =- \braket{S}{\Xi^0 n} = \frac{\gt}{\sqrt{40}}.
\end{aligned}
\end{equation}

\begin{table}[htb]
    \centering
    \begin{tabular}{|c|c|c|c|c|}
    \hline
             & SU(3) color & SU(3) flavor & SU(2) spin \\ \hline
        $S$  &  ${\tiny\Yvcentermath1 \yng(2,2,2)}=\mathbf{1}$  & ${\Yvcentermath1 \tiny\yng(2,2,2)}=\mathbf{1}$ & ${\Yvcentermath1 \tiny\yng(3,3)}=\mathbf{1}$ \\[1ex] \hline
        $B$  & ${\Yvcentermath1\tiny\yng(1,1,1)}=\mathbf{1}$ & ${\Yvcentermath1\tiny\yng(2,1)}=\mathbf{8}$ & ${\Yvcentermath1\tiny\yng(2,1)}=\mathbf{2}$ \\[1ex] \hline
    \end{tabular}
    \caption{Representations of $S$ and octet baryons under CSF.}
    \label{tab:young}
\end{table}

\section{Cooling of SN1987a}
\label{app:SN}
In this appendix, we recast the calculation in ref.~\cite{McDermott18} to constrain $\gt$ as a function of $m_S$. The hot and dense environment in proto-neutron stars can spontaneously create an abundant amount of hyperons. If the reaction rate of $\Lambda\Lambda \to S\gamma$ is rapid enough, $S$ would be brought into thermal equilibrium with baryons and hyperons in the star. The timescale needed for $S$ to equilibrate is estimated by ref.~\cite{McDermott18} to be
\begin{equation}
    t_S = \textrm{s}\, \frac{4\times10^{-34} \cmcs}{\expval{\sigma_{\Lambda\Lambda\to S\gamma}v}} \left(1+\frac{2\times10^{-32}\cmcs}{\expval{\sigma_{NN\to N\Lambda}v}}\right)^2,
\end{equation}
with
\begin{equation}
    \expval{\sigma_{NN\to N\Lambda}v} = 3\times10^{-30}\cmcs,\quad \expval{\sigma_{\Lambda\Lambda\to S\gamma}v}= 3\times10^{-23}\, \frac{g_{\Lambda S}^2\, (2m_\Lambda-m_S)}{176.9\textrm{MeV}} \cmcs.
\end{equation}
Note that the coupling $g_{\Lambda S}$ in their notation is related to $\gt$ by $\gt = \sqrt{40}g_{\Lambda S}$. Observations of neutrinos from SN1987a suggests a core-collapse supernovae from a proto-neutron star with a cooling time $\sim10$ s~\cite{Scholberg:2012id} (see, however, ref.~\cite{Bar19} for an alternate model). Requiring $t_S \gtrsim 10$ s implies the upper limit on $\gt$ shown as the red line in Figure~\ref{fig:exclusion}.


\section{\texorpdfstring{$S\to \Lambda p\pi$}{S to Lambda p pi}}
\label{app:belle}

\begin{figure}[htb]
    \centering
    \includegraphics[width=0.2\textwidth]{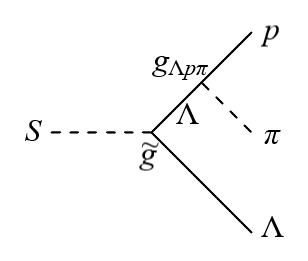}
    \caption{Feynman diagram for $S\to \Lambda p \pi$.}
    \label{fig:SLambdappi}
\end{figure}

Figure~\ref{fig:SLambdappi}
shows the effective Feynman diagram of the decay $S\to \Lambda p\pi$, and we construct the following meson-baryon Lagrangian
\begin{equation}
\label{eqn:SLambdappi}
    \mathcal{L} \supset  g_{\Lambda p\pi} \bar{\psi}_\Lambda (A+iB\gamma^5) \psi_p \phi_\pi + \frac{i \gt}{\sqrt{40}} \bar{\psi}_\Lambda \gamma^5 \psi_\Lambda^c \phi_S.
\end{equation}
The coefficients $A=1.05$ and $B=-7.15$ determine the strength of parity-violating and parity-conserving amplitudes of $\Lambda\to p\pi$~\cite{LambdaNPi}. To fix the value of $g_{\Lambda p\pi}$, we compute the decay rate
\begin{equation}
    \Gamma(\Lambda\to p\pi) = \frac{g_{\Lambda p \pi}^2}{8\pi}[A^2(E_p+m_p)+B^2(E_p-m_p)] \sqrt{\frac{(m_\Lambda^2+m_p^2-m_\pi^2)^2}{m_\Lambda^4}-\frac{4m_p^2}{m_\Lambda^2}} \, ,
\end{equation}
where $E_p= (m_\Lambda^2+m_p^2-m_\pi^2)/(2m_\Lambda)$ is the energy of the proton in the rest frame of $\Lambda$. By matching with the measured $\Lambda$ lifetime $ 2.6\times 10^{-10}$ s~\cite{PDG}, we find $g_{\Lambda p\pi}=3.87\times 10^{-7}$. Then, the decay rate of $S\to \Lambda p \pi$ can be evaluated by the integral
\begin{equation}
    \Gamma(S\to \Lambda p \pi) = \frac{1}{2m_S} \int d\Phi_3\, \langle |\mathcal{A}|^2 \rangle,
\end{equation}
where $\Phi_3$ is the 3-body phase space factor, and for simplicity we approximate $\langle|\mathcal{A}|^2 \rangle$ by a constant $\langle|\mathcal{A}|^2 \rangle \simeq {\gt^2} g_{\Lambda p\pi}^2/{40}$.

\bibliographystyle{JHEP}
\bibliography{main}

\providecommand{\noopsort}[1]{}\providecommand{\singleletter}[1]{#1}%

\providecommand{\href}[2]{#2}\begingroup\raggedright\begin{thebibliography}{10}

\bibitem{Farrar17}
G.~R. Farrar, \emph{{Stable Sexaquark}},
  \href{https://arxiv.org/abs/1708.08951}{{\ttfamily 1708.08951}}.

\bibitem{Farrar:2022mih}
G.~R. Farrar, \emph{{A Stable Sexaquark: Overview and Discovery Strategies}},
  \href{https://arxiv.org/abs/2201.01334}{{\ttfamily 2201.01334}}.

\bibitem{Jaffe77}
R.~L. Jaffe, \emph{{Perhaps a Stable Dihyperon}},
  \href{https://doi.org/10.1103/PhysRevLett.38.617,
  10.1103/PhysRevLett.38.195}{\emph{Phys. Rev. Lett.} {\bfseries 38} (1977)
  195}.

\bibitem{donoghue86}
J.~F. Donoghue, E.~Golowich and B.~R. Holstein, \emph{Weak decays of the h
  dibaryon}, \href{https://doi.org/10.1103/PhysRevD.34.3434}{\emph{Phys. Rev.
  D} {\bfseries 34} (1986) 3434}.

\bibitem{FarrarZaharijas}
G.~R. Farrar and G.~Zaharijas, \emph{{Nuclear and nucleon transitions of the H
  dibaryon}}, \href{https://doi.org/10.1103/PhysRevD.70.014008}{\emph{Phys.
  Rev.} {\bfseries D70} (2004) 014008}
  [\href{https://arxiv.org/abs/hep-ph/0308137}{{\ttfamily hep-ph/0308137}}].

\bibitem{hyper1991}
S.~Aoki et~al., \emph{{Direct observation of sequential weak decay of a double
  hypernucleus}}, \href{https://doi.org/10.1143/PTP.85.1287}{\emph{Prog. Theor.
  Phys.} {\bfseries 85} (1991) 1287}.

\bibitem{hyper20011}
J.~K. Ahn et~al., \emph{{Production of (Lambda Lambda)H-4 hypernuclei}},
  \href{https://doi.org/10.1103/PhysRevLett.87.132504}{\emph{Phys. Rev. Lett.}
  {\bfseries 87} (2001) 132504}.

\bibitem{hyper20012}
H.~Takahashi et~al., \emph{{Observation of a (Lambda Lambda)He-6 double
  hypernucleus}},
  \href{https://doi.org/10.1103/PhysRevLett.87.212502}{\emph{Phys. Rev. Lett.}
  {\bfseries 87} (2001) 212502}.

\bibitem{hyper2010}
{\scshape KEK-E176, J-PARC-E07} collaboration, \emph{{Double-Lambda hypernuclei
  via the Xi- hyperon capture at rest reaction in a hybrid emulsion}},
  \href{https://doi.org/10.1016/j.nuclphysa.2010.01.195}{\emph{Nucl. Phys.}
  {\bfseries A835} (2010) 207}.

\bibitem{SasakiHALQCD20}
{\scshape HAL QCD} collaboration, \emph{{$\Lambda\Lambda$ and N$\Xi$
  interactions from lattice QCD near the physical point}},
  \href{https://doi.org/10.1016/j.nuclphysa.2020.121737}{\emph{Nucl. Phys. A}
  {\bfseries 998} (2020) 121737}
  [\href{https://arxiv.org/abs/1912.08630}{{\ttfamily 1912.08630}}].

\bibitem{greenMainzH21}
J.~R. Green, A.~D. Hanlon, P.~M. Junnarkar and H.~Wittig, \emph{Weakly bound h
  dibaryon from su(3)-flavor-symmetric qcd},
  \href{https://doi.org/10.1103/physrevlett.127.242003}{\emph{Physical Review
  Letters} {\bfseries 127} (2021) }.

\bibitem{ALICE_LamLam19}
{\scshape ALICE} collaboration, \emph{{Study of the $\Lambda$-$\Lambda$
  interaction with femtoscopy correlations in pp and p-Pb collisions at the
  LHC}}, \href{https://doi.org/10.1016/j.physletb.2019.134822}{\emph{Phys.
  Lett. B} {\bfseries 797} (2019) 134822}
  [\href{https://arxiv.org/abs/1905.07209}{{\ttfamily 1905.07209}}].

\bibitem{Farrar18}
G.~R. Farrar, \emph{A precision test of the nature of dark matter and a probe
  of the qcd phase transition},
  \href{https://arxiv.org/abs/1805.03723}{{\ttfamily 1805.03723}}.

\bibitem{Strumia18}
C.~Gross, A.~Polosa, A.~Strumia, A.~Urbano and W.~Xue, \emph{{Dark Matter in
  the Standard Model?}},
  \href{https://doi.org/10.1103/PhysRevD.98.063005}{\emph{Phys. Rev.}
  {\bfseries D98} (2018) 063005}
  [\href{https://arxiv.org/abs/1803.10242}{{\ttfamily 1803.10242}}].

\bibitem{KolbTurner18}
E.~W. Kolb and M.~S. Turner, \emph{{Dibaryons cannot be the dark matter}},
  \href{https://doi.org/10.1103/PhysRevD.99.063519}{\emph{Phys. Rev.}
  {\bfseries D99} (2019) 063519}
  [\href{https://arxiv.org/abs/1809.06003}{{\ttfamily 1809.06003}}].

\bibitem{1815806}
N.~N. Achasov, J.~V. Bennett, A.~V. Kiselev, E.~A. Kozyrev and G.~N. Shestakov,
  \emph{{Evidence of the four-quark nature of $f_0$(980) and $f_0$(500)}},
  \href{https://doi.org/10.1103/PhysRevD.103.014010}{\emph{Phys. Rev. D}
  {\bfseries 103} (2021) 014010}
  [\href{https://arxiv.org/abs/2009.04191}{{\ttfamily 2009.04191}}].

\bibitem{farrarwintergerst}
G.~R. Farrar and N.~Wintergerst, \emph{{Wave Function of the Sexaquark or
  Compact H-dibaryon (In preparation)}}, 2023.

\bibitem{fwx20}
G.~R. Farrar, Z.~Wang and X.~Xu, \emph{{Dark Matter Particle in QCD}},
  \href{https://arxiv.org/abs/2007.10378}{{\ttfamily 2007.10378}}.

\bibitem{farrarxu}
X.~Xu and G.~R. Farrar, \emph{{Resonant scattering between dark matter and
  baryons: Revised direct detection and CMB limits}},
  \href{https://doi.org/10.1103/PhysRevD.107.095028}{\emph{Phys. Rev. D}
  {\bfseries 107} (2023) 095028}
  [\href{https://arxiv.org/abs/2101.00142}{{\ttfamily 2101.00142}}].

\bibitem{Rosner:1985yh}
J.~L. Rosner, \emph{{SU(3) Breaking and the $H$ Dibaryon}},
  \href{https://doi.org/10.1103/PhysRevD.33.2043}{\emph{Phys. Rev.} {\bfseries
  D33} (1986) 2043}.

\bibitem{Gignoux:1987cn}
C.~Gignoux, B.~Silvestre-Brac and J.~M. Richard, \emph{{Possibility of Stable
  Multi - Quark Baryons}},
  \href{https://doi.org/10.1016/0370-2693(87)91244-5}{\emph{Phys. Lett.}
  {\bfseries B193} (1987) 323}.

\bibitem{Callan:1985hy}
C.~G. Callan, Jr. and I.~R. Klebanov, \emph{{Bound State Approach to
  Strangeness in the Skyrme Model}},
  \href{https://doi.org/10.1016/0550-3213(85)90292-5}{\emph{Nucl. Phys.}
  {\bfseries B262} (1985) 365}.

\bibitem{Straub:1988mz}
U.~Straub, Z.-Y. Zhang, K.~Brauer, A.~Faessler and S.~B. Khadkikar,
  \emph{{Binding Energy of the Dihyperon in the Quark Cluster Model}},
  \href{https://doi.org/10.1016/0370-2693(88)90763-0}{\emph{Phys. Lett.}
  {\bfseries B200} (1988) 241}.

\bibitem{Nishikawa:1991di}
K.~Nishikawa, N.~Aoki and H.~Hyuga, \emph{{Hyperons and the H particle in the
  color dielectric model}},
  \href{https://doi.org/10.1016/0375-9474(91)90462-F}{\emph{Nucl. Phys.}
  {\bfseries A534} (1991) 573}.

\bibitem{Oka:1990vx}
M.~Oka and S.~Takeuchi, \emph{{Instanton Induced Interaction and the Strange
  Dibaryons}}, \href{https://doi.org/10.1016/0375-9474(91)90267-A}{\emph{Nucl.
  Phys.} {\bfseries A524} (1991) 649}.

\bibitem{Haidenbauer:2011za}
J.~Haidenbauer and U.~G. Meissner, \emph{{Exotic bound states of two baryons in
  light of chiral effective field theory}},
  \href{https://doi.org/10.1016/j.nuclphysa.2012.01.021}{\emph{Nucl. Phys.}
  {\bfseries A881} (2012) 44}
  [\href{https://arxiv.org/abs/1111.4069}{{\ttfamily 1111.4069}}].

\bibitem{LQCD11}
{\scshape HAL QCD} collaboration, \emph{{Bound H-dibaryon in Flavor SU(3) Limit
  of Lattice QCD}},
  \href{https://doi.org/10.1103/PhysRevLett.106.162002}{\emph{Phys. Rev. Lett.}
  {\bfseries 106} (2011) 162002}
  [\href{https://arxiv.org/abs/1012.5928}{{\ttfamily 1012.5928}}].

\bibitem{LQCD13}
{\scshape NPLQCD} collaboration, \emph{{Light Nuclei and Hypernuclei from
  Quantum Chromodynamics in the Limit of SU(3) Flavor Symmetry}},
  \href{https://doi.org/10.1103/PhysRevD.87.034506}{\emph{Phys. Rev.}
  {\bfseries D87} (2013) 034506}
  [\href{https://arxiv.org/abs/1206.5219}{{\ttfamily 1206.5219}}].

\bibitem{Kodama:1994np}
N.~Kodama, M.~Oka and T.~Hatsuda, \emph{{H dibaryon in the QCD sum rule}},
  \href{https://doi.org/10.1016/0375-9474(94)90908-3}{\emph{Nucl. Phys. A}
  {\bfseries 580} (1994) 445}
  [\href{https://arxiv.org/abs/hep-ph/9404221}{{\ttfamily hep-ph/9404221}}].

\bibitem{Azizi19}
K.~Azizi, S.~S. Agaev and H.~Sundu, \emph{{The Scalar Hexaquark $uuddss$: a
  Candidate to Dark Matter?}},
  \href{https://doi.org/10.1088/1361-6471/ab9a0e}{\emph{J. Phys. G} {\bfseries
  47} (2020) 095001} [\href{https://arxiv.org/abs/1904.09913}{{\ttfamily
  1904.09913}}].

\bibitem{Evans:2023zde}
N.~Evans and M.~Ward, \emph{{Running anomalous dimensions in holographic QCD:
  From the proton to the sexaquark}},
  \href{https://doi.org/10.1103/PhysRevD.108.026018}{\emph{Phys. Rev. D}
  {\bfseries 108} (2023) 026018}
  [\href{https://arxiv.org/abs/2304.10816}{{\ttfamily 2304.10816}}].

\bibitem{Buccella}
F.~Buccella, \emph{{On the Mass and on the Dynamical Properties of the
  Sexaquark}}, \href{https://doi.org/10.22323/1.376.0024}{\emph{PoS} {\bfseries
  CORFU2019} (2020) 024}.

\bibitem{mackeen18}
D.~McKeen, A.~E. Nelson, S.~Reddy and D.~Zhou, \emph{{Neutron stars exclude
  light dark baryons}},
  \href{https://doi.org/10.1103/PhysRevLett.121.061802}{\emph{Phys. Rev. Lett.}
  {\bfseries 121} (2018) 061802}
  [\href{https://arxiv.org/abs/1802.08244}{{\ttfamily 1802.08244}}].

\bibitem{Baym:2018ljz}
G.~Baym, D.~Beck, P.~Geltenbort and J.~Shelton, \emph{{Testing dark decays of
  baryons in neutron stars}},
  \href{https://doi.org/10.1103/PhysRevLett.121.061801}{\emph{Phys. Rev. Lett.}
  {\bfseries 121} (2018) 061801}
  [\href{https://arxiv.org/abs/1802.08282}{{\ttfamily 1802.08282}}].

\bibitem{shahrbaf+22}
M.~Shahrbaf, D.~Blaschke, S.~Typel, G.~R. Farrar and D.~E. Alvarez-Castillo,
  \emph{{Sexaquark dilemma in neutron stars and its solution by quark
  deconfinement}},
  \href{https://doi.org/10.1103/PhysRevD.105.103005}{\emph{Phys. Rev. D}
  {\bfseries 105} (2022) 103005}
  [\href{https://arxiv.org/abs/2202.00652}{{\ttfamily 2202.00652}}].

\bibitem{Poulin:2016nat}
V.~Poulin, P.~D. Serpico and J.~Lesgourgues, \emph{{A fresh look at linear
  cosmological constraints on a decaying dark matter component}},
  \href{https://doi.org/10.1088/1475-7516/2016/08/036}{\emph{JCAP} {\bfseries
  08} (2016) 036} [\href{https://arxiv.org/abs/1606.02073}{{\ttfamily
  1606.02073}}].

\bibitem{McDermott18}
S.~D. McDermott, S.~Reddy and S.~Sen, \emph{{Deeply bound dibaryon is
  incompatible with neutron stars and supernovae}},
  \href{https://doi.org/10.1103/PhysRevD.99.035013}{\emph{Phys. Rev.}
  {\bfseries D99} (2019) 035013}
  [\href{https://arxiv.org/abs/1809.06765}{{\ttfamily 1809.06765}}].

\bibitem{SlatyerWu16}
T.~R. Slatyer and C.-L. Wu, \emph{{General Constraints on Dark Matter Decay
  from the Cosmic Microwave Background}},
  \href{https://doi.org/10.1103/PhysRevD.95.023010}{\emph{Phys. Rev.}
  {\bfseries D95} (2017) 023010}
  [\href{https://arxiv.org/abs/1610.06933}{{\ttfamily 1610.06933}}].

\bibitem{Liu:2020wqz}
H.~Liu, W.~Qin, G.~W. Ridgway and T.~R. Slatyer,
  \emph{{Lyman-\ensuremath{\alpha} constraints on cosmic heating from dark
  matter annihilation and decay}},
  \href{https://doi.org/10.1103/PhysRevD.104.043514}{\emph{Phys. Rev. D}
  {\bfseries 104} (2021) 043514}
  [\href{https://arxiv.org/abs/2008.01084}{{\ttfamily 2008.01084}}].

\bibitem{WW21}
D.~Wadekar and Z.~Wang, \emph{{Strong constraints on decay and annihilation of
  dark matter from heating of gas-rich dwarf galaxies}},
  \href{https://doi.org/10.1103/PhysRevD.106.075007}{\emph{Phys. Rev. D}
  {\bfseries 106} (2022) 075007}
  [\href{https://arxiv.org/abs/2111.08025}{{\ttfamily 2111.08025}}].

\bibitem{deuteron}
M.~Garcon and J.~W. Van~Orden, \emph{{The Deuteron: Structure and
  form-factors}}, \href{https://doi.org/10.1007/0-306-47915-X_4}{\emph{Adv.
  Nucl. Phys.} {\bfseries 26} (2001) 293}
  [\href{https://arxiv.org/abs/nucl-th/0102049}{{\ttfamily nucl-th/0102049}}].

\bibitem{SNO}
{\scshape SNO} collaboration, \emph{{The Sudbury Neutrino Observatory}},
  \href{https://doi.org/10.1016/j.nuclphysb.2016.04.035}{\emph{Nucl. Phys.}
  {\bfseries B908} (2016) 30}
  [\href{https://arxiv.org/abs/1602.02469}{{\ttfamily 1602.02469}}].

\bibitem{Siegert:2015knp}
T.~Siegert, R.~Diehl, G.~Khachatryan, M.~G. Krause, F.~Guglielmetti, J.~Greiner
  et~al., \emph{{Gamma-ray spectroscopy of Positron Annihilation in the Milky
  Way}}, \href{https://doi.org/10.1051/0004-6361/201527510}{\emph{Astron.
  Astrophys.} {\bfseries 586} (2016) A84}
  [\href{https://arxiv.org/abs/1512.00325}{{\ttfamily 1512.00325}}].

\bibitem{DeRocco:2019fjq}
W.~DeRocco and P.~W. Graham, \emph{{Constraining Primordial Black Hole
  Abundance with the Galactic 511 keV Line}},
  \href{https://doi.org/10.1103/PhysRevLett.123.251102}{\emph{Phys. Rev. Lett.}
  {\bfseries 123} (2019) 251102}
  [\href{https://arxiv.org/abs/1906.07740}{{\ttfamily 1906.07740}}].

\bibitem{superk}
{\scshape Super-Kamiokande} collaboration, \emph{{Search for Nucleon and
  Dinucleon Decays with an Invisible Particle and a Charged Lepton in the Final
  State at the Super-Kamiokande Experiment}},
  \href{https://doi.org/10.1103/PhysRevLett.115.121803}{\emph{Phys. Rev. Lett.}
  {\bfseries 115} (2015) 121803}
  [\href{https://arxiv.org/abs/1508.05530}{{\ttfamily 1508.05530}}].

\bibitem{BNV}
J.~Heeck and V.~Takhistov, \emph{{Inclusive Nucleon Decay Searches as a
  Frontier of Baryon Number Violation}},
  \href{https://doi.org/10.1103/PhysRevD.101.015005}{\emph{Phys. Rev. D}
  {\bfseries 101} (2020) 015005}
  [\href{https://arxiv.org/abs/1910.07647}{{\ttfamily 1910.07647}}].

\bibitem{Peccei:1977hh}
R.~D. Peccei and H.~R. Quinn, \emph{{CP Conservation in the Presence of
  Instantons}}, \href{https://doi.org/10.1103/PhysRevLett.38.1440}{\emph{Phys.
  Rev. Lett.} {\bfseries 38} (1977) 1440}.

\bibitem{Dodelson:1993je}
S.~Dodelson and L.~M. Widrow, \emph{{Sterile-neutrinos as dark matter}},
  \href{https://doi.org/10.1103/PhysRevLett.72.17}{\emph{Phys. Rev. Lett.}
  {\bfseries 72} (1994) 17}
  [\href{https://arxiv.org/abs/hep-ph/9303287}{{\ttfamily hep-ph/9303287}}].

\bibitem{Berezhiani:2015yta}
Z.~Berezhiani, A.~D. Dolgov and I.~I. Tkachev, \emph{{Reconciling Planck
  results with low redshift astronomical measurements}},
  \href{https://doi.org/10.1103/PhysRevD.92.061303}{\emph{Phys. Rev. D}
  {\bfseries 92} (2015) 061303}
  [\href{https://arxiv.org/abs/1505.03644}{{\ttfamily 1505.03644}}].

\bibitem{Enqvist:2015ara}
K.~Enqvist, S.~Nadathur, T.~Sekiguchi and T.~Takahashi, \emph{{Decaying dark
  matter and the tension in $\sigma_8$}},
  \href{https://doi.org/10.1088/1475-7516/2015/09/067}{\emph{JCAP} {\bfseries
  09} (2015) 067} [\href{https://arxiv.org/abs/1505.05511}{{\ttfamily
  1505.05511}}].

\bibitem{FrancoAbellan:2020xnr}
G.~Franco~Abell\'an, R.~Murgia, V.~Poulin and J.~Lavalle, \emph{{Implications
  of the $S_8$ tension for decaying dark matter with warm decay products}},
  \href{https://doi.org/10.1103/PhysRevD.105.063525}{\emph{Phys. Rev. D}
  {\bfseries 105} (2022) 063525}
  [\href{https://arxiv.org/abs/2008.09615}{{\ttfamily 2008.09615}}].

\bibitem{essigetal13}
R.~Essig, E.~Kuflik, S.~D. McDermott, T.~Volansky and K.~M. Zurek,
  \emph{{Constraining Light Dark Matter with Diffuse X-Ray and Gamma-Ray
  Observations}}, \href{https://doi.org/10.1007/JHEP11(2013)193}{\emph{JHEP}
  {\bfseries 11} (2013) 193} [\href{https://arxiv.org/abs/1309.4091}{{\ttfamily
  1309.4091}}].

\bibitem{localdensity}
J.~I. Read, \emph{{The Local Dark Matter Density}},
  \href{https://doi.org/10.1088/0954-3899/41/6/063101}{\emph{J. Phys.}
  {\bfseries G41} (2014) 063101}
  [\href{https://arxiv.org/abs/1404.1938}{{\ttfamily 1404.1938}}].

\bibitem{NFM}
D.~A. Neufeld, G.~R. Farrar and C.~F. McKee, \emph{{Dark Matter that Interacts
  with Baryons: Density Distribution within the Earth and New Constraints on
  the Interaction Cross-section}},
  \href{https://doi.org/10.3847/1538-4357/aad6a4}{\emph{Astrophys. J.}
  {\bfseries 866} (2018) 111}
  [\href{https://arxiv.org/abs/1805.08794}{{\ttfamily 1805.08794}}].

\bibitem{Cooke:2013cba}
R.~Cooke, M.~Pettini, R.~A. Jorgenson, M.~T. Murphy and C.~C. Steidel,
  \emph{{Precision measures of the primordial abundance of deuterium}},
  \href{https://doi.org/10.1088/0004-637X/781/1/31}{\emph{Astrophys. J.}
  {\bfseries 781} (2014) 31} [\href{https://arxiv.org/abs/1308.3240}{{\ttfamily
  1308.3240}}].

\bibitem{Coc:2015bhi}
A.~Coc, P.~Petitjean, J.-P. Uzan, E.~Vangioni, P.~Descouvemont, C.~Iliadis
  et~al., \emph{{New reaction rates for improved primordial D/H calculation and
  the cosmic evolution of deuterium}},
  \href{https://doi.org/10.1103/PhysRevD.92.123526}{\emph{Phys. Rev. D}
  {\bfseries 92} (2015) 123526}
  [\href{https://arxiv.org/abs/1511.03843}{{\ttfamily 1511.03843}}].

\bibitem{belle}
{\scshape Belle} collaboration, \emph{{Search for an $H$-dibaryon with mass
  near $2m_\Lambda$ in $\Upsilon(1S)$ and $\Upsilon(2S)$ decays}},
  \href{https://doi.org/10.1103/PhysRevLett.110.222002}{\emph{Phys. Rev. Lett.}
  {\bfseries 110} (2013) 222002}
  [\href{https://arxiv.org/abs/1302.4028}{{\ttfamily 1302.4028}}].

\bibitem{kek1}
{\scshape KEK-PS E224} collaboration, \emph{{Enhanced Lambda Lambda production
  near threshold in the C-12(K-,K+) reaction}},
  \href{https://doi.org/10.1016/S0370-2693(98)01416-6}{\emph{Phys. Lett.}
  {\bfseries B444} (1998) 267}.

\bibitem{kek2}
C.~J. Yoon et~al., \emph{{Search for the H-dibaryon resonance in C-12 (K-, K+
  Lambda Lambda X)}},
  \href{https://doi.org/10.1103/PhysRevC.75.022201}{\emph{Phys. Rev.}
  {\bfseries C75} (2007) 022201}.

\bibitem{BNL1}
{\scshape BNL-E888} collaboration, \emph{{Search for diffractive dissociation
  of a longlived H dibaryon}},
  \href{https://doi.org/10.1103/PhysRevD.53.R3487}{\emph{Phys. Rev.} {\bfseries
  D53} (1996) R3487}.

\bibitem{BNL2}
{\scshape BNL-E888} collaboration, \emph{{Search for the weak decay of an H
  dibaryon}}, \href{https://doi.org/10.1103/PhysRevC.56.1164,
  10.1103/PhysRevLett.76.3277}{\emph{Phys. Rev. Lett.} {\bfseries 76} (1996)
  3277} [\href{https://arxiv.org/abs/hep-ex/9603002}{{\ttfamily
  hep-ex/9603002}}].

\bibitem{BNLE836:1997}
{\scshape BNL E836} collaboration, \emph{{Search for H dibaryon in He-3 (K-,
  k+) Hn}}, \href{https://doi.org/10.1103/PhysRevLett.78.3646}{\emph{Phys. Rev.
  Lett.} {\bfseries 78} (1997) 3646}.

\bibitem{AertsDoverPRL82}
A.~T.~M. Aerts and C.~B. Dover, \emph{{THE (K-, K+) REACTION AND THE H
  DIBARYON}}, \href{https://doi.org/10.1103/PhysRevLett.49.1752}{\emph{Phys.
  Rev. Lett.} {\bfseries 49} (1982) 1752}.

\bibitem{AertsDoverPRD83}
A.~t.~m. Aerts and C.~b. Dover, \emph{{ON THE PRODUCTION OF THE SIX QUARK H
  DIBARYON IN THE (K-, K+) REACTION}},
  \href{https://doi.org/10.1103/PhysRevD.28.450}{\emph{Phys. Rev. D} {\bfseries
  28} (1983) 450}.

\bibitem{Bar19}
N.~Bar, K.~Blum and G.~D'Amico, \emph{{Is there a supernova bound on axions?}},
  \href{https://doi.org/10.1103/PhysRevD.101.123025}{\emph{Phys. Rev. D}
  {\bfseries 101} (2020) 123025}
  [\href{https://arxiv.org/abs/1907.05020}{{\ttfamily 1907.05020}}].

\bibitem{Isgur:1978wd}
N.~Isgur and G.~Karl, \emph{{Positive Parity Excited Baryons in a Quark Model
  with Hyperfine Interactions}},
  \href{https://doi.org/10.1103/PhysRevD.19.2653}{\emph{Phys. Rev. D}
  {\bfseries 19} (1979) 2653}.

\bibitem{farrarwang23}
G.~R. Farrar and Z.~Wang, \emph{In preparation},  2023.

\bibitem{hardcore}
R.~Wiringa, \emph{Private communication}, 2019.

\bibitem{Naghdi:2007ek}
M.~Naghdi, \emph{{Nucleon-nucleon interaction: A typical/concise review}},
  \href{https://doi.org/10.1134/S1063779614050050}{\emph{Phys. Part. Nucl.}
  {\bfseries 45} (2014) 924}
  [\href{https://arxiv.org/abs/nucl-th/0702078}{{\ttfamily nucl-th/0702078}}].

\bibitem{HAMADA1962382}
T.~Hamada and I.~Johnston, \emph{A potential model representation of
  two-nucleon data below 315 mev},
  \href{https://doi.org/https://doi.org/10.1016/0029-5582(62)90228-6}{\emph{Nuclear
  Physics} {\bfseries 34} (1962) 382}.

\bibitem{Reid1968}
J.~{Reid}, Roderick~V., \emph{{Local phenomenological nucleon-nucleon
  potentials}},
  \href{https://doi.org/10.1016/0003-4916(68)90126-7}{\emph{Annals of Physics}
  {\bfseries 50} (1968) 411}.

\bibitem{Lonardoni:2017egu}
D.~Lonardoni, A.~Lovato, S.~C. Pieper and R.~B. Wiringa, \emph{{Variational
  calculation of the ground state of closed-shell nuclei up to $A=40$}},
  \href{https://doi.org/10.1103/PhysRevC.96.024326}{\emph{Phys. Rev. C}
  {\bfseries 96} (2017) 024326}
  [\href{https://arxiv.org/abs/1705.04337}{{\ttfamily 1705.04337}}].

\bibitem{Shahrbaf:2022upc}
M.~Shahrbaf, D.~Blaschke, S.~Typel, G.~R. Farrar and D.~E. Alvarez-Castillo,
  \emph{{Sexaquark dilemma in neutron stars and its solution by quark
  deconfinement}},
  \href{https://doi.org/10.1103/PhysRevD.105.103005}{\emph{Phys. Rev. D}
  {\bfseries 105} (2022) 103005}
  [\href{https://arxiv.org/abs/2202.00652}{{\ttfamily 2202.00652}}].

\bibitem{BringmannPospelovCRDM19}
T.~Bringmann and M.~Pospelov, \emph{{Novel direct detection constraints on
  light dark matter}},
  \href{https://doi.org/10.1103/PhysRevLett.122.171801}{\emph{Phys. Rev. Lett.}
  {\bfseries 122} (2019) 171801}
  [\href{https://arxiv.org/abs/1810.10543}{{\ttfamily 1810.10543}}].

\bibitem{Alvey:2022pad}
J.~Alvey, T.~Bringmann and H.~Kolesova, \emph{{No room to hide: implications of
  cosmic-ray upscattering for GeV-scale dark matter}},
  \href{https://doi.org/10.1007/JHEP01(2023)123}{\emph{JHEP} {\bfseries 01}
  (2023) 123} [\href{https://arxiv.org/abs/2209.03360}{{\ttfamily
  2209.03360}}].

\bibitem{fx23}
G.~R. Farrar and X.~Xu, \emph{{Limits on moderately-interacting Cosmic Ray Dark
  Matter in the GeV mass range in underground detectors}},
  \href{https://arxiv.org/abs/in preparation}{{\ttfamily in preparation}}.

\bibitem{cannoni}
M.~Cannoni, \emph{{Relativistic $<\sigma v_\text{rel}>$ in the calculation of
  relics abundances: a closer look}},
  \href{https://doi.org/10.1103/PhysRevD.89.103533}{\emph{Phys. Rev.}
  {\bfseries D89} (2014) 103533}
  [\href{https://arxiv.org/abs/1311.4508}{{\ttfamily 1311.4508}}].

\bibitem{thomsom}
M.~Thomson, \emph{Mordern Particle Physics}. Cambridge University Press, 2013.

\bibitem{McNAMEE}
P.~McNAMEE, S.~J. and F.~CHILTON, \emph{Tables of clebsch-gordan coefficients
  of s${\mathrm{u}}_{3}$},
  \href{https://doi.org/10.1103/RevModPhys.36.1005}{\emph{Rev. Mod. Phys.}
  {\bfseries 36} (1964) 1005}.

\bibitem{Scholberg:2012id}
K.~Scholberg, \emph{{Supernova Neutrino Detection}},
  \href{https://doi.org/10.1146/annurev-nucl-102711-095006}{\emph{Ann. Rev.
  Nucl. Part. Sci.} {\bfseries 62} (2012) 81}
  [\href{https://arxiv.org/abs/1205.6003}{{\ttfamily 1205.6003}}].

\bibitem{LambdaNPi}
A.~Parreno, C.~Bennhold and B.~R. Holstein, \emph{{An EFT for the weak Lambda N
  interaction}},
  \href{https://doi.org/10.1016/j.nuclphysa.2005.01.008}{\emph{Nucl. Phys.}
  {\bfseries A754} (2005) 127}
  [\href{https://arxiv.org/abs/nucl-th/0312047}{{\ttfamily nucl-th/0312047}}].

\bibitem{PDG}
{\scshape Particle Data Group} collaboration, \emph{{Review of Particle
  Physics}}, \href{https://doi.org/10.1103/PhysRevD.98.030001}{\emph{Phys.
  Rev.} {\bfseries D98} (2018) 030001}.

\end{thebibliography}\endgroup

\end{document}